\documentclass[twoside,11pt]{article}

\usepackage{times}
\usepackage{graphicx}
\usepackage{amsfonts}
\usepackage{amssymb}
\usepackage{amsmath} 
\usepackage{amsthm}
\usepackage{natbib}
\usepackage{graphicx}

\usepackage[margin=0.8in]{geometry}

\newcommand {\ignore}[1]{}               

\newcommand {\br}[1]{\left(#1\right)}

\newcommand {\cbr}[1]{\left\{#1 \right\}}

\newcommand {\abs}[1]{\left\vert\, #1 \,\right\vert}

\newcommand {\mc}[1]{\mathcal{#1}}

\newcommand{\RR}{\mathbb{R}}    

\newcommand{\Hcal}{\mathcal{H}} 

\newcommand{\Xcal}{\mathcal{X}} 

\newcommand{\II}{\mbox{\boldmath $1$}}

\bibpunct{(}{)}{;}{a}{,}{,}

\title{Virtual screening with support vector machines and structure kernels}

\author{Pierre Mah\'e  \\
       Xerox Research Center Europe\\
       6, chemin de Maupertuis, 38240 Meylan, France\\
        pierre.mahe@xrce.xerox.com \\
       \and Jean-Philippe Vert  \\
       Centre for Computational Biology, Ecole des Mines de Paris\\
       35, rue Saint Honor\'e, 77305 Fontainebleau, France \\
        jean-philippe.vert@ensmp.fr }

\begin{document}

\maketitle
\abstract{Support vector machines and kernel methods have recently gained considerable attention in chemoinformatics. They offer generally good performance for problems of supervised classification or regression, and provide a flexible and computationally efficient framework to include relevant information and prior knowledge about the data and problems to be handled. In particular, with kernel methods molecules do not need to be represented and stored explicitly as vectors or fingerprints, but only to be \emph{compared} to each other through a comparison function technically called a \emph{kernel}. While classical kernels can be used to compare vector or fingerprint representations of molecules, completely new kernels were developed in the recent years to directly compare the 2D or 3D structures of molecules, without the need for an explicit vectorization step through the extraction of molecular descriptors. While still in their infancy, these approaches have already demonstrated their relevance on several toxicity prediction and structure-activity relationship problems.}

\section*{Introduction}

Computational approaches play an increasingly important role in modern drug discovery. In particular, accurate predictive models accounting for the biological activity and drug-likeliness of candidate molecules can help in the identification of promising molecules and screening for various side-effects, leading to substantial savings in terms of time and costs for the development of new drugs. Such predictive models aim at inferring a relationship between the structure of a molecule and its biological and chemical properties, including toxicity, pharmacocinetics and activity against a target. The development of high-throughput technologies to assay such properties for large numbers of candidate molecules, and the subsequent availability of increasing quantities of molecules with characterized properties, has triggered the use of statistical and machine learning approaches to automatically \emph{learn} the structure-property relationship from these pools of characterized molecules.

Decades of research in machine learning and statistics have provided a profusion of methods for that purpose, ranging from classical least-square linear regression to artificial neural networks or decision trees \citep{Hastie2001elements}. While each method has is specificities, strengths and weaknesses, a common issue when one wants to infer a structure-property relationship concerns the way molecules are represented. While small molecules are often represented as 2D or 3D structures in chemistry and chemoinformatics, most statistical methods, including linear models and nonlinear neural networks, require vectors an input. Molecules must therefore be first encapsulated as finite-dimensional vectors, using various molecular descriptors, before being presented as input to these algorithms. The construction of molecular descriptors is however a difficult task. Often a significant chemical expertise coupled with heuristic feature selection methods is needed to chose, among the plethora of possible molecular descriptors, the most relevant ones for a property to be predicted. The number of molecular descriptors must moreover be kept as small as possible to limit the complexity of the inference task.

An alternative to this issue has emerged recently with the advent of support vector machines (SVM) and related kernel methods in machine learning \citep{Vapnik1998Statistical,Scholkopf2002Learning,Shawe-Taylor2004Kernel}. SVM is an algorithm for pattern recognition and regression that provides a useful framework to overcome the difficulty of data representations as vectors of low dimensions, both from a theoretical and a computational point of view. Theoretically, first, SVM are able to infer models in large or even infinite dimensions from a finite number of observations. Indeed the complexity of the learning task is not directly related to the dimension of the input vectors, but rather to some measure of complexity of the classification rules which are precisely controlled by SVM through the use of regularization \citep{Vapnik1998Statistical}. Practically, second, a computational trick known as the \emph{kernel trick} allows the estimation of models with a complexity that does not depend on the dimension of the input, but only on the number of training points. Hence training a model with vectors of infinite dimension is no more computationally demanding than training a model for small fingerprints -- as long as the so-called \emph{kernel function}, which corresponds to the inner product of the vectors, can be computed efficiently. Combined together these properties give SVM the ability to work with molecules represented by vectors of large or even infinite dimension in a computationally efficient framework, leveraging the burden of feature selection and giving the modelers new opportunities to imagine large sets of molecular descriptors.

SVM often provide state-of-the-art performances on many classification and regression tasks, and enjoy therefore an increasing popularity in various application fields, including bioinformatics and chemoinformatics \citep{Schoelkopf2004Kernel}. For example, SVM have been applied to the prediction of the activity of molecules on a number of target classes \citep{Burbidge2001Drug,Weston2003Feature, Arimoto2005Development, Briem2005Classifying, Liu2004QSAR, Saeh2005Lead, Tobita2005discriminant}, toxicological properties \citep{Kramer2002Fragment, Helma2004Data, Luan2005Classification}, drug-likeliness \citep{Byvatov2003Comparison, Mueller2005Classifying, Takaoka2003Development}, blood-brain barrier permeability \citep{Doniger2002Predicting}, enantioselectivity \citep{Aires-de-Sousa2005Prediction}, aqueous solubility \citep{Lind2003Support}, or isoelectric point \citep{Liu2004Prediction}, to name just a few.

While most recent successful applications of SVM in chemoinformatics were obtained by just plugging classical molecular descriptors to the SVM, an increasing line of work seeks to investigate the unique opportunities offered by SVM to go beyond classical fingerprints and molecular descriptors, thanks to the kernel trick. This avenue was pioneered simultaneously and independently by \citet{Kashima2003Marginalized} and \citet{Gartner2003graph} who proposed to represent the 2D structure of a molecule by an infinite-dimensional vector of linear fragment counts and showed how SVM can handle this representation with the kernel trick. Later work quickly refined these 2D kernels \citep{Kashima2004Kernels,Mahe2005Graph,Ralaivola2005Graph} and proposed new infinite-dimensional representations of 3D structures \citep{Swamidass2005Kernels,Mahe2006Pharmacophore,Azencott2007One-}. 

These first attempts to enlarge the flexibility of molecular descriptor-based predictive models represent a promising direction for \emph{in silico} modelling of structure-property relationship, because they illustrate the unique possibilities offered by SVM and more generally kernel methods in this context. We review them in this paper with the hope to offer a state-of-the-art description of the latest development in this field, and an invitation for the chemoinformatics community to further investigate these possibilities. For that purpose we first provide a quick introduction to SVM and kernels in Section \ref{sec:svm}, and illustrate the relevance of the kernel trick when working with 2D structures of molecules with a simple example of 2D kernel in Section \ref{sec:2d-kernel}. This example is further generalized and connected to recent work on 2D kernels in Section \ref{sec:2d-extensions}, and practical issues with these kernels are discussed in Section \ref{sec:2d-practice}. In Section \ref{sec:3d} we present another approach that focuses on the representation of 3D structures of molecules, and discuss practical issues for this approach in Section \ref{sec:3d-practice}. We conclude by a discussion and suggestions for future work in Section \ref{sec:discussion}.

\section{Support vector machines and kernels}\label{sec:svm}

SVM is a machine learning algorithm for pattern recognition originally developed in the early 1990's by V. Vapnik and coworkers \citep{Boser1992training,Vapnik1998Statistical}. Although various extensions to multiclass classification, regression, outlier detection or feature construction also exist, we focus in this review on the simple pattern recognition problem and refer the interested reader to various textbooks to know more about these extensions, collectively known as \emph{kernel methods} \citep{Scholkopf2002Learning,Shawe-Taylor2004Kernel}. A pattern recognition problem occurs when one is given a finite set of objects that belong to two possible classes, and must \emph{learn} from this training set a rule to automatically \emph{predict} the class of objects with unknown class. This general and abstract formulation encompasses in fact a number of practical situations in chemoinformatics and beyond. We focus here in particular on situations where the objects available are small molecules, and the classes to be predicted represents various properties of interest such as toxic/non toxic, druggable/non-druggable, or inhibitor/non-inhibitor of a given target. Hence a typical pattern recognition problem could be, given a list of toxic and non-toxic molecules, to learn a rule to predict whether a new candidate molecule is toxic or not.

\begin{figure}
\begin{center}
\includegraphics[width=0.5\textwidth]{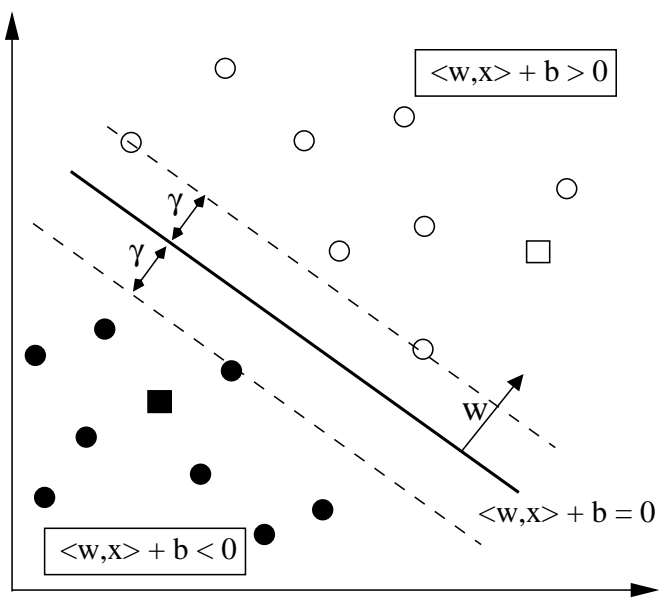}
\caption{SVM estimates a linear separation between the classes. When the training patterns are linearly separable and the trade-off parameter $C$ in Equation (\ref{eq:primal}) is set to $+\infty$, then the separating hyperplane selected by SVM is the one that maximizes the distance to the closest point on each side ($\gamma$ on this picture). In general, some training points may be misclassified by the selected hyperplane to control overfitting.}\label{fig:hyperplane}
\end{center}
\end{figure}

More formally, we represent the training set available as a set of $n$ objects $x_{1},\ldots,x_{n} \in \mathcal{X}$, where $\Xcal$ denotes the set of all possible objects, and associated binary labels $y_1,\ldots,y_{n} \in \{ -1 ; 1 \}$. In our case each object $x_{i}$ represents a molecule, $\Xcal$ denotes the set of all possible molecules, and the two classes $1$ and $-1$ are arbitrary representations of two classes or interest, such as ``toxic'' and ``non-toxic''. Pattern recognition algorithms, such as SVM, use this training set to produce a classifier $f : \mathcal{X} \mapsto \{ -1 ; 1 \}$ that can be used to predict the class of any new data $x \in  \mathcal{X}$ by the value $f(x)$. When objects are $d$-dimensional vectors, that is, $\mathcal{X} = \mathbb{R}^d$, the classifier output by SVM is based on the sign of a linear function: 
\begin{equation}\label{eq:f}
f(x) = \text{sign}( \langle w,x \rangle + b)\,,
\end{equation} 
for some $(w,b) \in \mathcal{X} \times \mathbb{R}$ defined below. In this case the classifier has a geometric interpretation: the hyperplane $\langle w,x \rangle + b=0$ separates the input space $\mathcal{X}$ into two half-spaces, and the prediction of the class of a new point depends on its position on the one or on the other side of the hyperplane. The particular hyperplane selected by SVM is the one that solves the following optimization problem :
\begin{equation}\label{eq:primal}
\underset{w,b}{\min} \cbr{\frac{1}{2} ||w||^2  + C \sum_{i=1}^n L\left( y_i , \langle w,x_i   \rangle +b \right)} ,
\end{equation}  
where $C$ is a parameter and $L(y,t)$ is the \emph{hinge loss} function equal to $0$ if $yt\geq 1$, and $1-yt$ otherwise. For a given training example $(x_i,y_i)$, the hinge loss term $L\left( y_i , \langle w,x_i   \rangle +b \right)$ quantifies how ``good'' the prediction $\langle w,x_i   \rangle +b$ of a candidate classifier $(w,b)$ is, in the sense that the better the prediction, the smaller the loss. For example there is no loss when $w$ and $b$ are such that $y_{i}\br{\langle w,x_i   \rangle +b} \geq 1$, which means that $\langle w,x_i   \rangle +b$ has the sign of $y_{i}$ and is larger than $1$ in absolute value. In other words the loss is zero when the prediction is correct and made with large confidence. Now the second term in the sum (\ref{eq:primal}) is the average loss over the training set of the candidate classifier $(w,b)$: it is small when the classifier fits well the training points, i.e., makes on average ``good'' predictions. On the other hand, the first term $||w||^2$ in (\ref{eq:primal}) is small when the slope of the classifier is small. The two terms in (\ref{eq:primal}) are often in conflicts, especially in large dimension, because it is often difficult to fit the training points well with linear functions of limited slopes. The rational behind the optimization problem (\ref{eq:primal}) is indeed to find a linear classifier that reaches a trade-off between the goodness of fit on the training set (as quantified by the second term of this sum), and the smoothness of the classifier (as quantified by the first term). The parameter $C$ controls this trade-off, by balancing the importance of each term. In the extreme case when $C=+\infty$ and the training points can be correctly separated by a hyperplane, then no error is allowed on the training set and the classifier with largest margin is found (Figure \ref{fig:hyperplane}).

It is often interesting to rewrite problem (\ref{eq:primal}) in an equivalent way, using classical optimization theory. Indeed, this problem is equivalent to the following quadratic problem, called its dual:
\begin{equation}\label{eq:dual}
\begin{split}
        & \underset{\alpha\in\RR^n}{\max} \cbr{\sum_{i=1}^n \alpha_i - \frac{1}{4} \sum_{i,j=1}^{n} \alpha_i \alpha_j y_i y_j \langle x_i , x_j \rangle }\,,\\
        & \text{subject to : } \sum_{i=1}^{n} \alpha_i y_i = 0 \;  \text{and } 0 \leq \alpha_i \leq C \; , \; i \in [1:n]\,.
\end{split}
\end{equation}  
Both problems (\ref{eq:primal}) and (\ref{eq:dual}) are equivalent in the sense that the solution $(w^*,b^*)$ of the primal problem (\ref{eq:primal}) can be deduced from the solution $\alpha^*$ of the dual problem (\ref{eq:dual}). In particular, it can be shown that  $w^* = \sum_{i=1}^n \alpha_i^* y_i x_i $, and $b^*$ can also be deduced from $\alpha^*$. As a result, the decision function (\ref{eq:f}) can also be expressed in terms of the solution $\alpha^*$ of the dual problem:
\begin{equation}\label{eq:classif}
f(x) = \text{sign} \br{ \sum_{i=1}^n \alpha_i^* \langle x , x_i \rangle + b^* }\,.
\end{equation}

\begin{figure}
\begin{center}
\includegraphics[width=0.75\textwidth]{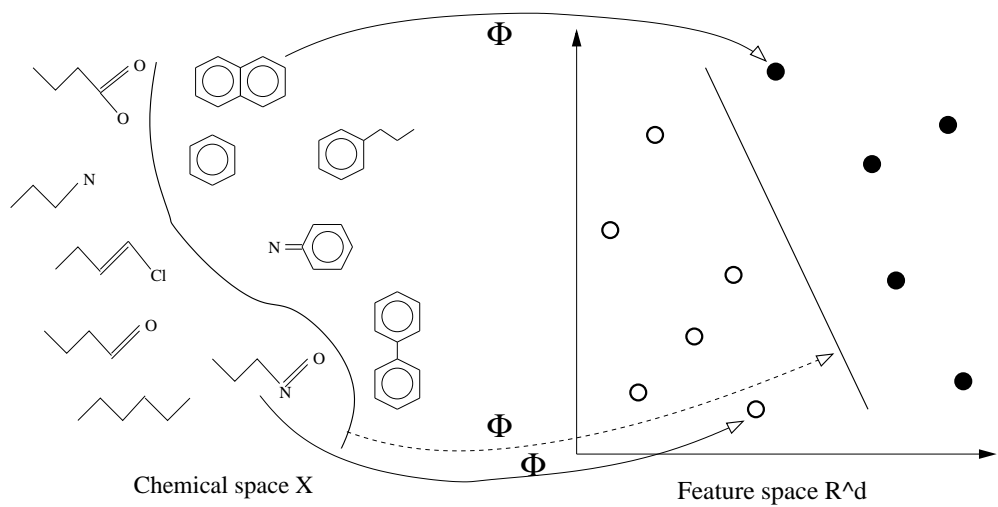}
\caption{In order to use SVM with molecules, we need to define an embedding of the space of molecules to a vector space, i.e., a representation of each molecule $x$ as a vector $\Phi(x)$. Note that, contrary to usual fingerprint-based approaches, the vector space might have a large or even infinite dimension.}\label{fig:mapping}
\end{center}
\end{figure}

Let us now consider the use SVM for pattern recognition with molecules, represented for example by their 2D or 3D structures. Such structures being not vectors, they can not be directly input to SVM. Instead we need to embed the set of 2D or 3D structures of molecules $\Xcal$ to a vector space $\Hcal$ through a mapping $\Phi : \Xcal \rightarrow \Hcal$. We can then apply the SVM algorithm to the training vectors $\Phi(x_{i}), i=1,\ldots,n$, as illustrated in Figure \ref{fig:mapping}.
An important point to notice is that in the dual formulation (\ref{eq:dual}), the data are only present through dot-products: pairwise dot-products between the training points during the learning phase in (\ref{eq:dual}), and dot-products between a new data and the training points during the prediction phase in (\ref{eq:classif}). This means that instead of explicitly knowing $\Phi(x)$ for any $x \in \Xcal$, it suffices to be able to compute inner products of the form:
\begin{equation}\label{eq:k}
k(x,x')=\langle \Phi(x) , \Phi(x')\rangle\,,
\end{equation}
for any $x,x' \in \Xcal$. In that case the dual optimization problem (\ref{eq:dual}) solved by SVM can be rewritten as follows:
\begin{equation}\label{eq:dualkernel}
\begin{split}
        & \underset{\alpha\in\RR^n}{\max} \cbr{ \sum_{i=1}^n \alpha_i - \frac{1}{4} \sum_{i,j=1}^{n} \alpha_i \alpha_j y_i y_j k( x_i , x_j ) }\,,\\
        & \text{subject to : } \sum_{i=1}^{n} \alpha_i y_i = 0 \;  \text{and } 0 \leq \alpha_i \leq C \; , \; i \in [1:n]\,.
\end{split}
\end{equation}
Moreover the classification function (\ref{eq:f}) becomes:
\begin{equation}\label{eq:fkernel}
f(x) = \text{sign} \br{ \sum_{i=1}^n \alpha_i^* k(x , x_i) + b^* }\,.
\end{equation}
Hence we see that for both the training of the SVM (\ref{eq:dualkernel}) and the prediction of the class of new points ($\ref{eq:fkernel}$), the feature map $\Phi$ only appears through the function $k$, which is called a \emph{kernel}.
Importantly it is sometimes easier to compute directly the kernel $k(x,x')$ between two points than their explicit representations as vectors in $\Hcal$. In fact a classical result of \citet{Aronszajn1950Theory} characterizes all functions $k:\Xcal\times\Xcal\mapsto\RR$ that are valid kernel, i.e., for which there exists a feature space $\Hcal$ and a mapping $\Phi : \Xcal \rightarrow \Hcal$ such that (\ref{eq:k}) holds (they constitute the so-called class of \emph{positive definite} functions). Hence, with this characterization at hand, any kernel $k$ can be used with a SVM as long as it satisfies the positive definiteness property. 

The formulation of SVM in terms of kernels (\ref{eq:dualkernel}-\ref{eq:fkernel}) offers at least two major advantages over the formulation in terms of explicit vectors (\ref{eq:dual}-\ref{eq:classif}). First, it enables the straightforward extension of the linear SVM to non linear decision functions by using a nonlinear kernel, while keeping its nice properties intact (e.g., unicity of the solution, robustness to over-fitting, etc...). As an example, the Gaussian kernel $k(x,x') = \exp\br{-||x-x'||^2 / 2\sigma^2}$ is positive definite and can therefore be used as a kernel in the SVM algorithm (\ref{eq:dualkernel}). Plugging this kernel into (\ref{eq:fkernel}) we see that the resulting discrimination function has the form:
$$
f(x) = \text{sign} \cbr{\sum_{i=1}^n \alpha^*_i \exp\br{-\frac{||x-x_{i}||^2 }{ 2\sigma^2}}}\,,
$$
which is clearly a nonlinear function of $x$. Second, this formulation offers the possibility to directly apply SVM to non-vectorial data, such as 2D or 3D structures of molecules, provided a positive definite kernel function to compare these structures is defined. The definition of such \emph{structure kernels} for molecules is explained in the following sections.

\section{A simple kernel for 2D structures}\label{sec:2d-kernel}

It is common to describe the 2D structure of a molecule as a labeled undirected graph $G=(V,E)$, with atoms as vertices $V$ and covalent bounds as edges $E$. Here we assume that a label is assigned to each node and edge, typically to describe the type of atoms and bounds involved. In order to train linear models for structure-property relationship prediction, each labeled graph $G$ representing a molecule must first be transformed into a vector $\Phi(G)$. In this section we describe a simple vector representation obtained by counting all walks of a given length $n$, and show the relevance of the kernel formulation in this case.

\begin{figure}
\begin{center}
\includegraphics[width=0.75\textwidth]{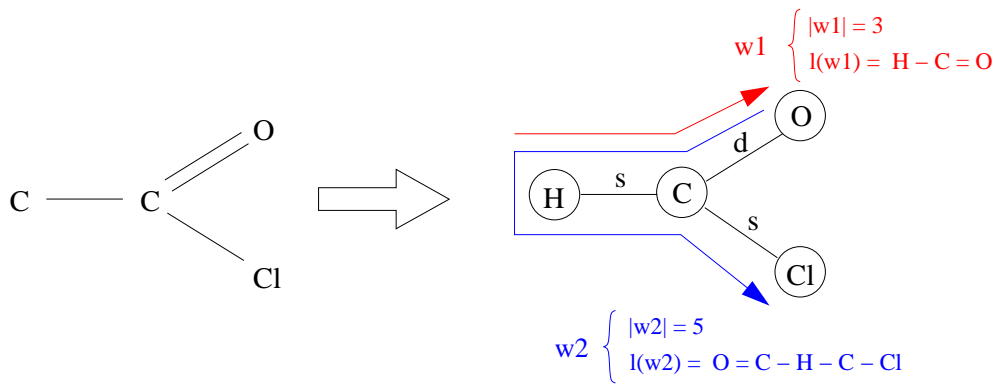}
\caption{The 2D structure of a molecule (on the left) can be represented by a labeled graph (on the right). Two walks on the graph are illustrated, together with their label and length.}\label{fig:walks}
\end{center}
\end{figure}

A walk of length $n$ on a graph is a sequence of $n$ adjacent vertices. We note that this definition allows a given vertex or edge to be present more than once in a walk. Clearly, the number of walks of length $n$ on a graph $G$ is finite, and we denote by $W_{n}(G)$ this set of walks in the following. By concatenating the labels of the vertices and of the edges of a walk $w$ we obtain a sequence of labels which we denote by $l(w)$, the label of the walk $w$. Moreover, we note $L_{n}$ the set of possible labels for walks of length $n$, i.e., all possible sequences alternating $n$ vertex labels with $n-1$ edge labels. Figure \ref{fig:walks} illustrates these definitions. Now a simple way to represent a graph $G$ by a vector is to extract all walks of length $n$ from its structure, sort them by label, and count in $\Phi(G)$ the number of walks with each possible label in $L_{n}$. In other words the dimension of $\Phi(G)$ is equal to the size of $L_{n}$, and for each possible walk label $l\in L_{n}$ we define the coordinate $\Phi_{l}(G)$ as the number of walks in $G$ having label $l$. More formally the feature $\Phi_{l}(G)$ is defined by:
\begin{equation}\label{eq:featvector}
\Phi_{l}(G) = \sum_{w\in W_{n}(G)} \II(l(w) = l)\,.
\end{equation}

A direct approach to train a linear model with these vector representations would require the explicit computation and storage of $\Phi(G)$ for all graphs $G$ in a dataset. This approach becomes problematic when $n$ becomes large, because the number of walk labels increases exponentially with $n$. As an example, keeping only $6$ types of atoms and $3$ types of covalent bounds, the number of possible labels reaches $1,944$ for walks of length $3$; $34,992$ for walks of length $4$; $629,856$ for walks of length $5$; and more than $3$ billions for walks of length $8$. This explosion in the dimension  of $\Phi(G)$ suggests in practice either to restrict oneself to walks of length 2 or 3, or to compress the representation $\Phi(G)$. The later approach is widely used in chemoinformatics because fragments of length $5$-$10$ are known to provide useful information in many structure-property relationship problems. The solution most often encountered is to use a hash table of limited size (typically $1024$ or $2048$) to map the vector $\Phi(G)$ onto a vector of smaller dimension, called a molecular fingerprint \citep{Gasteiger2003Chemoinformatics}. An obvious drawback of this solution is the danger of clashes, i.e., the mapping of different labels to the same position in the hashed vector.

\begin{figure}
\begin{center}
\includegraphics[width=0.75\textwidth]{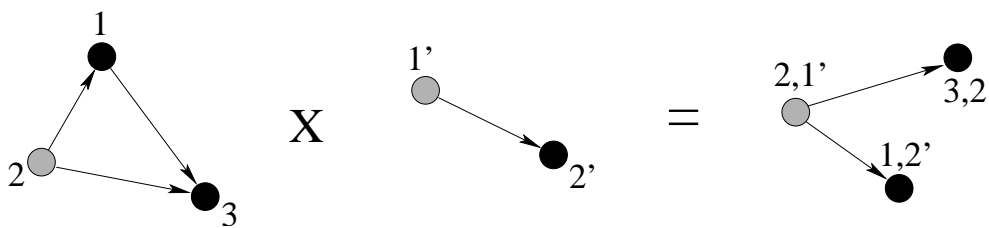}
\caption{The product graph of two graphs (on the left) is obtained by considering all pairs of vertices with similar labels as vertices, and connecting two such vertices when the respective pairs of vertices in the initial graphs are connected (on the right). Each walk in the product graph (e.g., $(2,1')-(3,2')$) is associated to a pair of walks in the initial graphs with same labels (e.g., $2-3$ and $1'-2'$), and vice-versa.}\label{fig:product-graph}
\end{center}
\end{figure}

An alternative solution for the use of large $n$ values is to use kernels. As we now show, indeed, kernels allow the estimation of linear models for vectors $\Phi(G)$ without reducing their dimension nor requiring the computation and storage of the vectors. Indeed, remembering from Section \ref{sec:svm} that SVM only need the definition of the inner product between vectors to estimate a linear problem, we only need to show how the inner product for the vector representation (\ref{eq:featvector}) can be computed efficiently. For that purpose, let us write this inner product more explicitly for any two graphs $G$ and $G'$:
\begin{equation}\label{eq:trick}
\begin{split}
\langle \Phi(G) , \Phi(G')\rangle
& = \sum_{l\in L_{n}} \Phi_{l}(G) \Phi_{l}(G') \\
& = \sum_{l\in L_{n}} \br{\sum_{w\in W_{n}(G)} \II(l(w) = l)} \br{\sum_{w'\in W_{n}(G')} \II(l(w') = l)} \\
& = \sum_{w\in W_{n}(G)} \sum_{w'\in W_{n}(G')} \br{\sum_{l\in L_{n}} \II(l(w) = l) \II(l(w') = l)} \\
& = \sum_{w\in W_{n}(G)} \sum_{w'\in W_{n}(G')} \II(l(w)=l(w'))\,.
\end{split}
\end{equation}
In other words the inner product between $\Phi(G)$ and $\Phi(G')$ can be expressed exactly as the number of pairs of walks $(w,w')$ of length $n$, respectively in $G$ and $G'$, with the same label. In order to show how this number can be computed efficiently, it is useful to introduce the \emph{product graph} $G\times G'$ which is a graph whose vertices are pairs of vertices of $G$ and $G'$ with the same label, and whose edges connect pairs of vertices which are connected both in $G$ and $G'$ (Figure \ref{fig:product-graph}). In other words the vertices of $G\times G'$ are the pairs $(v,v')\in V\times V'$ with $l(v)=l(v')$, and there is an edge between $(v_{1},v_{1}')$ and $(v_{2},v_{2}')$ if and only if there is both an edge between $v_{1}$ and $v_{2}$ in $G$ and an edge between $v_{1}'$ and $v_{2}'$ in $G'$, and if both edges have the same label. It is easy to see, then, that a walk in the product graph is a sequence of pairs of vertices $(v,v')$, in $G$ and $G'$, that are connected in $G\times G'$ and therefore in $G$ and $G'$. Moreover both sequences of vertices in $G$ and $G'$ are made of pairs of vertices and pairs of edges with the same label, i.e., they form a pair of walks in $G$ and $G'$ with the same label. Conversely, given any walks $w$ in $G$ and $w'$ in $G'$ with same label $l(w)=l(w')$, there is a walk in the product graph that corresponds to the pair of walks $(w,w')$. In other words, there is a bijection between the pairs of walks in $G$ and $G'$ with the same label, on the one hand, and the walks on $G\times G'$, on the other hand. Hence counting the number of pairs of walks of length $n$ on $G$ and $G'$ with the same label is equivalent to simply counting the number of walks of length $n$ on $G\times G'$, as illustrated in Figure \ref{fig:product-graph}. It turns out that counting the number of walks of length $n$ on a general graph (and in particular on a product graph for our purpose) can be easily computed by a recursion over $n$. Indeed, for a general graph, if we denote by $A_{i}(v)$ the number of walks of length $i$ starting at vertex $v$, then $A_{1}(u) = 1$ for any vertex $u$ and the following recursion formula holds:
\begin{equation}\label{eq:recursion}
A_{i+1}(v) = \sum_{u\sim v} A_{i}(u)\,,
\end{equation}
where the sum is over the neighbor vertices of $v$. $A_{n}(u)$ can therefore be computed for any $u\in V$ by applying this formula recursively over $i$. The number of walks of length $n$ on the graph is then simply obtained by summing $A_{n}(u)$ over the vertices $u$. We observe that if we denote by $A$ the adjacency matrix of the graph and by $\II$ the vector whose entries are all equal to $1$, then (\ref{eq:recursion}) simply expresses $A^{i+1}\II$ as $A \times A^{i}\II$, and the count of walks of length $n$ is equal to $\II^\top A^{n-1} \II$. 

To summarize, we have shown that for the vector representation (\ref{eq:featvector}), the inner product between two graphs $G$ and $G'$ representing the 2D structures of two molecules can be computed by (i) constructing the adjacency matrix $A$ of the product graph $G\times G'$ and (ii) computing $\II^\top A^{n-1} \II$ using the recursion (\ref{eq:recursion}). This computation is exact and efficient, although the dimension of the vectors can reach the billions. In particular, the complexity of the computation increases only linearly with $n$, while the number of features increases exponentially. Using this inner product with a kernel method for pattern recognition or regression allows to estimate a linear model in this space without ever computing nor storing any vector.

\section{2D kernel extensions}\label{sec:2d-extensions}

The kernel for 2D structures presented in the previous section to illustrate the power of kernels can be used as such, but many extensions have been proposed to increase the flexibility and the expressiveness of the representation. In this section we review some of these extensions.

\subsection{Walks of various lengths}\label{sec:infinite}
In the computation of the kernel based on walks of length $n$, we note that kernels of length $i<n$ are computed as intermediaries. The choice of $n$ is arbitrary in practice and should depend on the targeted application and the data available. Alternatively we may decide not to choose a particular value of $n$, but to combine walks of different lengths in a joint feature model. The inner product being additive when new features are added, the kernel corresponding to the feature space (\ref{eq:featvector}) where all walks of length up to $n$ are considered is the sum of the kernels corresponding walks of fixed length smaller than $n$. The complexity of the computation is barely increased for this extension: instead of performing the recursion (\ref{eq:recursion}) $n$ times before summing the terms, one just need to increase a counter by the sum of the terms at each iteration.

When $n$ increases the inner product in this "until-$n$" extension grows exponentially with $n$ and diverges. A solution if one wishes to use large values for $n$, and even infinite $n$ to be able to include all walks, is to weight the contributions of different walks by a factor $\lambda(w)$ that will ensure convergence of the series, i.e., to consider the following kernel:
\begin{equation}
k(G,G') = \sum_{n=1}^\infty \sum_{w\in W_{n}(G)} \sum_{w'\in W_{n}(G')} \lambda(w)\lambda(w') \II(l(w)=l(w'))\,.
\end{equation}
As an example, \citet{Gartner2002Exponential} proposed to weight the contribution of walks of length $i$ in the inner product by a factor $\beta^{i/2}$, i.e., to consider the formula:
\begin{equation}\label{eq:geometric}
\begin{split}
k(G,G') &= \sum_{n=1}^\infty \sum_{w\in W_{n}(G)} \sum_{w'\in W_{n}(G')} \beta^n \II(l(w)=l(w'))\\
&=\sum_{n=1}^\infty \beta^{n} k_{n}(G,G')\,,
\end{split}
\end{equation}
where $k_{n}$ denotes the kernel based on the count of walks of length exactly $n$. Remembering from the previous Section that $k_{n}(G,G')$ is equal to $\II^\top A^{n-1} \II$, where $A$ is the adjacency matrix of $G\times G'$, we can rewrite and factorize this kernel as follows for $\beta$ small enough:
\begin{equation}
\begin{split}
k(G,G') &= \sum_{n=1}^\infty \beta^{n} \II^\top A^{n-1} \II \\
&=  \beta \II^\top \br{\sum_{n=0}^\infty \beta^{n} A^{n} } \II \\
&= \beta \II^\top \br{I-\beta A}^{-1} \II\,.
\end{split}
\end{equation}
Hence the computation of the inner product in the infinite-dimensional space of all walk counts can be performed explicitly, at the cost of inverting the sparse matrix $I-\beta A$. In practice the first terms of the power series expansion provide a fast and good approximation to the complete kernel, and allow more flexibility in the weighting of the walks of different length.

Another weighting scheme for walks has been proposed independently by \citet{Kashima2003Marginalized}, who propose to define Markov random walks of each graph and weight the occurrence of each walk on a graph by its probability under the corresponding random walk model. As for the exponential decay, the random walk weighting scheme factorizes along the walks and can be computed with the same tricks as the exponential decay walk kernel.

\subsection{Filtering tottering walks}\label{sec:no-totters}
In the previous section, we did not make any restriction on the definition of walks: they are simply defined as successions of connected graph vertices.
Because molecular graphs are essentially undirected, this generic definition allows walks to have an erratic behaviour, which can lead in turn to a misleading information about the true structure of the graph in the kernel.
Indeed, arbitrarily long walks can for instance be generated by simply alternating between two connected vertices.
A natural way to increase the expressive power of walks with respect to the structure of the graphs is to prevent vertices from appearing more than once in a walk.
In the terminology of graph theory, this corresponds to defining a kernel based on common {\em paths} instead of common walks.
Albeit very natural, this extension unfortunately renders the kernel computation untractable, as pointed out by \citet{Gartner2003graph}.

\begin{figure}
\begin{center}
\includegraphics[width=0.4\textwidth]{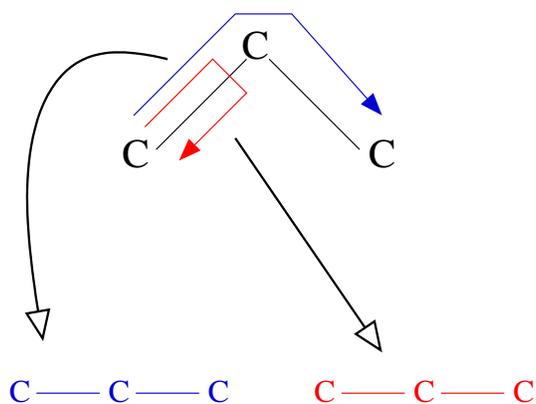}
\caption{Tottering (red) and no-tottering (blue) walks.  These two walks are labeled as a succession of 3 carbon atoms, but only the blue one involves 3 distinct atoms.}\label{fig:tottering}
\end{center}
\end{figure}

A computationally efficient alternative proposed by \citet{Mahe2005Graph} is to disregard the \emph{tottering walks} in the enumeration of walks. As illustrated in Figure \ref{fig:tottering}, a tottering walk is a walk that comes back to a vertex it has just left. Although the notion of path is stronger than the notion of tottering walks for general graphs, they are equivalent on graphs without cycles. The relevance of the concept of tottering walks stems from computational advantages: as shown by \citet{Mahe2005Graph} the set of tottering walks of a graph $G$ corresponds to a set of walks of a transformed graph $t(G)$, where the transformation $t$ involves adding additional vertices and edges. As a result, the kernel for two graphs $G$ and $G'$ based on non-tottering walks only is easily computed as the standard walk kernel between the transformed graphs $t(G)$ and $t(G')$. More details about this transformation can be found in \citet{Mahe2005Graph}.

\subsection{Increasing the expressiveness of walks}\label{sec:morgan}
A second criticism that can be made to walk kernels is the fact that, because of their linearity, walks bear limited information about the structure of a graph. A principled way to address this issue, which is actually the topic of the next subsection, is to introduce subgraphs of a higher level of complexity in the kernel construction. 
In practice, however, this approach usually raises additional complexity issues that can be hard to circumvent.
A simpler alternative is to keep a walk-based characterization of graphs and introduce some form of prior knowledge in the graph labeling function, in order to enrich the information brought by walks about the graph structure.
This is in particular the approach taken in \citet{Mahe2005Graph} where a new set of labels is defined for the vertices of a graph, based on the local environment of the atoms in the corresponding molecule.
This method relies on a topological index, called the {\em Morgan index}, which is defined for each atom of the molecule according to the following iterative procedure.
Initially, the index associated to every vertex is equal to 1. Then, at each step, the index of a vertex is defined as the sum of the indices associated to its neighbors at the previous iteration.
This process is straightforward to implement in practice, since if we let $M_i$ be the vector of Morgan indices computed at the $i$-th iteration, it reads as $M_0 = {\bf 1}$ and $M_{i+1} = A M_i$, where ${\bf 1}$ is the unity vector and $A$ is the adjacency matrix of the graph.

\begin{figure}
\begin{center}
\includegraphics[width=\textwidth]{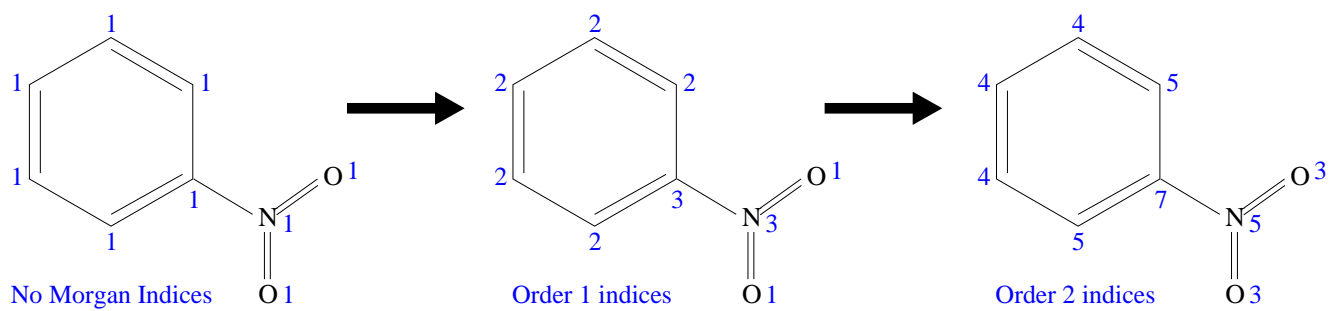}
\caption{Illustration of the Morgan process. Initially, all atoms of the cycle are seen as identical. For increasing iterations, the presence of the \texttt{NO2} branch is more and more reflected in the atoms of the cycle.}\label{fig:morgan}
\end{center}
\end{figure}

As illustrated in Figure \ref{fig:morgan}, Morgan indices make it possible to distinguish between atoms having the same type but different topological properties.
When they are included in the labels of the vertices, these indices therefore define a walk as a sequence of atoms taken in a particular topological configuration.
In practice, the advantage of this refinement is twofold.
First, the introduction of topological information in walk labels enriches the information they bear with respect to the structure of the graphs to be compared.
Second, because atoms are made more specific to the graph they belong to, as illustrated in Figure \ref{fig:morgan}, the number of identically labeled atoms found in a pair of graphs automatically decreases, which has the effect of reducing the size of their product graph, hence the time of computing the kernel. Note that this computation advantage is surprisingly due to the increase in dimension of the feature space.
We note however that while this Morgan process systematically reduces the cost of computing the kernel, performing too many iterations makes it impossible to detect common walks within a pair of graphs.

\subsection{Subtree kernels}
\begin{figure}[h]
        \begin{center}
        \includegraphics[width=0.75\textwidth]{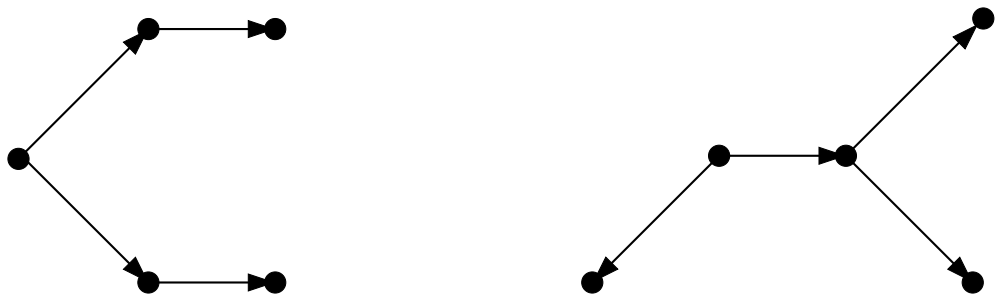}
        \caption{Two graphs having the same walk content, namely $\bullet : \times 5$ ; $\bullet \!\!\! \rightarrow \!\!\! \bullet : \times 4$ and  $\bullet \!\! \rightarrow \!\!\! \bullet \!\!\! \rightarrow \!\!\! \bullet : \times 2 $, and consequently mapped to the same point of the feature space corresponding a kernel based on the count of walks \citep{Gartner2003graph}. \label{fig:tree-identical} }
        \end{center}
\end{figure}

As mentioned in the previous subsection, the linear nature of walks limits their ability to properly encode the structure of a graph.
This fact is emphasized by \citet{Ramon2003Expressivity} who show that graphs can be structurally different yet have the same walk content, which makes them indistinguishable by a kernel based on the count of common walks.
Figure \ref{fig:tree-identical} illustrates this issue on a simple example.
On the other hand, they also show that computing a perfect graph kernel, that is, a kernel mapping non-isomorphic graphs to distinct points in the feature space, is at least as hard as solving the graph isomorphism problem for which there is no known polynomial-time algorithm. This suggests that the expressiveness of graph kernels must be traded for their computational complexity.

\begin{figure}
\begin{center}
\includegraphics[width=0.75\textwidth]{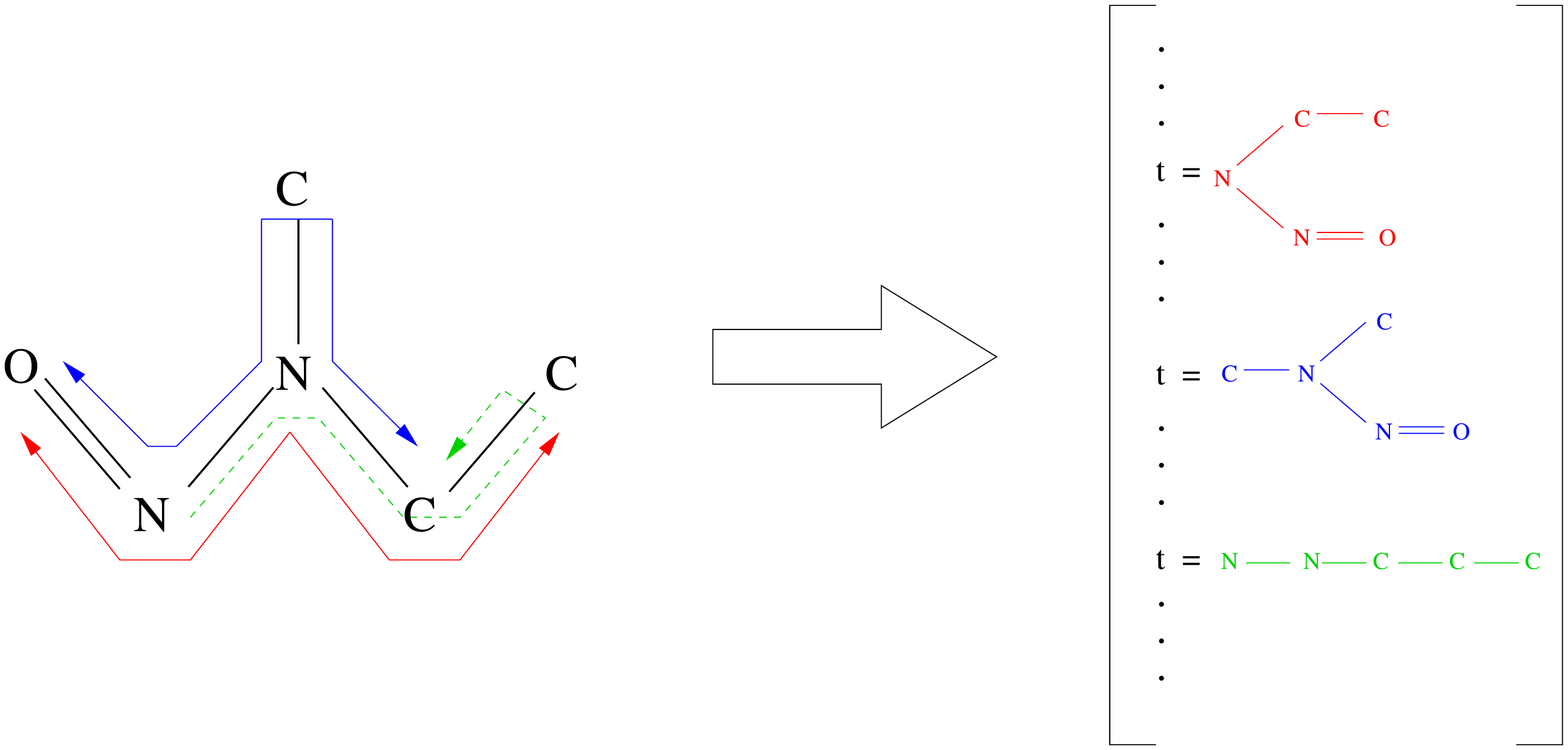}
\caption{Illustration of the tree-structured fragment representation: a graph $G$ (left) and an extract of its feature space representation $\phi(G)$ (right). Note that the green tree corresponds to a walk structure.}\label{fig:subtree}
\end{center}
\end{figure}

As a first step towards a refinement of the feature space used in walk-based graph kernels, \cite{Ramon2003Expressivity} introduce a kernel function comparing graphs on the basis of their common subtrees. 
As illustrated in Figure \ref{fig:subtree}, this representation looks particularly promising for molecules, since it allows to capture in a principled way a wide range of functional features of molecules, that typically correspond to specific branching patterns on their associated graphs.
On the practical side, this type of kernels can be computed by means of dynamic programming algorithms that recursively detect and extend identical neighborhood properties within the vertices of the graphs to be compared, in order to explicitly build their set of common subtrees.
The relative contribution of subtrees of different sizes is typically controlled by means of a parameter playing a similar role to that of the parameter $\beta$ in Equation (\ref{eq:geometric}).
These algorithms have a prohibitive complexity in general, but they can be deployed for molecular graphs where, because of valence rules, the degree of the vertices is small in average.
The relevance of this class of kernels, as well as its relationship with standard walk-based kernels, has been analyzed in details in \cite{Mahe2006Graph}.

\section{2D kernels in practice}\label{sec:2d-practice}
As a conclusion about kernels for 2D structures, we now discuss several issues related to their application in practice.

        \subsection{Implementation and complexity issues}\label{sec:2d-implementation}
As mentioned in Section \ref{sec:2d-kernel} an elegant way to compute walk based kernels lies in the product graph formalism initially introduced by \citet{Gartner2003graph}. 
The basic idea of the product graph construction is to merge the pair of graphs to be compared into a single graph, in such a way that a bijection is defined between the set of walks of the product graph and the set of common walks of the two initial graphs.
It then follows that the number of walks of a given length occuring at the same time in the two graphs can be obtained by simple matrix products, which actually offers a closed form solution to the computation of kernels based on walks of infinite length for well chosen walk weighting schemes \citep{Gartner2003graph,Kashima2004Kernels}.
As a result, even though the dimensionality associated to these kernels can be very large, and actually infinite, computing these kernels under the product graph formalism has a polynomial complexity with respect to the product of the size of the graphs to be compared\footnote{More precisely, the worst case complexity is cubic with respect to the product of the size of the graphs.}.

In practice, this type of product-graph implementations remains time consuming, even for relatively small graphs, which questions the suitability of these kernels for virtual screening applications, that typically involve large datasets of molecules.
However, if only walks up to a given length are considered, which usually makes sense for real world applications, fast algorithms can be used to compute walk kernels, based for instance on trie tree structures and string kernel algorithms \citep{Leslie2002Spectrum,Shawe-Taylor2004Kernel}, or standard depth-first search procedures \citep{Ralaivola2005Graph}.
Moreover, alternative implementations allowing to drastically reduce the time needed to compute such kernels in their general form have recently been proposed\citep{Vishwanathan2007Fast}.

        \subsection{Kernel normalization}\label{sec:normalize}
A potential drawback of kernels comparing structured objects by means of their substructures lies in the fact that kernel values are highly dependent on the size of the objects to be compared.
Indeed, big objects tend to be granted a higher degree of similarity than small objects for the only reason that they are made of a larger number of substructures.
This fact can lead to a serious bias of the subsequent prediction model, and the classical way to tackle this issue it to apply a normalization operation in order to take into account the size of the objects in the value of the kernel function.
In practice, the mainstream normalization scheme is given by the following expression:
$$
\tilde{k}(x,y) = \frac{k(x,y)}{\sqrt{k(x,x)k(y,y)}},
$$
where $k$ is the original kernel, and $\tilde{k}$ its normalized value.
Note that this normalization operation has the effect of setting the diagonal of the kernel matrix to one, meaning that individual objects are given the same degree of self-similarity, whatever their size is. Geometrically, this amounts to scaling all vectors to unit norm before taking their inner product.

In the context of molecular graph kernels, alternative normalization schemes based on the Tanimoto similarity coefficient have recently been introduced \citep{Ralaivola2005Graph,Swamidass2005Kernels}.
The Tanimoto coefficient is widely used in chemoinformatics to assess the similarity of molecular fingerprints.
For a pair of fingerprints $(A,B)$, it is defined as:
$$
T_{AB} = \frac{A^{\top}B}{A^{\top}A + B^{\top}B - A^{\top}B}.
$$
For binary fingerprints, it can be seen as the ratio between their intersection, that is, the number of bits set to one in both fingerprints, and their union.
As pointed out by \citet{Ralaivola2005Graph}, since it is based on inner product operations, this coefficient can be generalized to any kernel function, leading to the notion of {\em Tanimoto kernel}, defined for a kernel $k$ as:
$$
\tilde{k}(x,y) = \frac{k(x,y)}{k(x,x) + k(y,y) - k(x,y)}.
$$
This transformation provides an alternative way to normalize kernel functions in the sense that $\tilde{k}(x,x) = 1$ for all $x$.
Several variations on this idea, that allow to generalize the classical Tanimoto coefficient in different ways, are proposed in \citep{Ralaivola2005Graph,Swamidass2005Kernels}.

        \subsection{Kernel parameterization}
Last but not least comes the issue of kernel parameterization.
This question is of tremendous importance since a bad parameterization can seriously entail the success of the subsequent virtual screening application.
First one must choose to consider the kernel based on walks or the kernel based on subtrees. As for now, this questions remains largely open since apart from the study of \citet{Mahe2006Graph}, the relevance of subtrees in graph kernels has not been studied in details.
While the preliminary results presented in this study suggest that subtree kernels may indeed improve over their walk-based counterparts, they also show that this class of substructures raises additional issues, related in particular to the computational complexity of the kernels as well as the explosion in the number of subtrees found in the graphs.

Concerning the parameterization of walk kernels, the main issue concerns the length(s) of the walks to consider: either walks of a precise length, up to a maximal (but finite) length, or even up to infinite length.
In practice, this question is highly dependent on the problem considered.
Optimally choosing this parameter can therefore hardly be made {\em a priori} but involves cross-validation procedures.
Although focusing on walks of a precise length can be optimal in some cases\footnote{For instance, walks of length 6 or 7 can be optimal to characterize molecules mainly made of aromatic cycles.}, a safe default choice is to consider walks of length up to a limited value to be taken around 8 or 10.
Actually, because kernels based on an infinite number of walks require to down-weight the contribution of walks depending on their length (as in Equation (\ref{eq:geometric}) for instance), long walks are in practice so penalized that their individual contribution is barely taken into account in the kernel.
Explicitly limiting the length of the walks to be taken into account therefore makes sense in practice.
Moreover, considering a finite number of walks provides a greater flexibility in the way to control their relative contribution in the kernel, and offers the practical advantage of paving the way to the deployment of computationally cheaper algorithms, as discussed in Section \ref{sec:2d-implementation}.
A second important issue is related to kernel normalization.
Although the impact of choosing the first or the second normalization scheme introduced in Section \ref{sec:normalize} has not been analysed in details, Tanimoto kernels led to good results in several validation studies \citep{Ralaivola2005Graph,Swamidass2005Kernels,Azencott2007One-}.
Finally, one may consider further refinements such as filtering tottering walks and introducing Morgan indices.
As shown in \citet{Mahe2005Graph}, Morgan indices of a limited order, typically obtained at the 2nd or 3rd iteration of the process, can indeed improve virtual screening models while reducing their computational costs\footnote{Actually, this is only true for product-graph implementations. For trie-tree implementations, Morgan indices have the opposite effect of increasing the cost of computing the kernel.}.
Filtering tottering walks should be subject to caution however.
Indeed, as shown in \citet{Mahe2006Graph}, while this can indeed improve the models in some cases, it seems that the tottering phenomenon can also be helpful to detect similarity between structurally different compounds.

\section{A 3D pharmacophore kernel}\label{sec:3d}
\begin{figure*}
\centering
\includegraphics[width=0.24\textwidth]{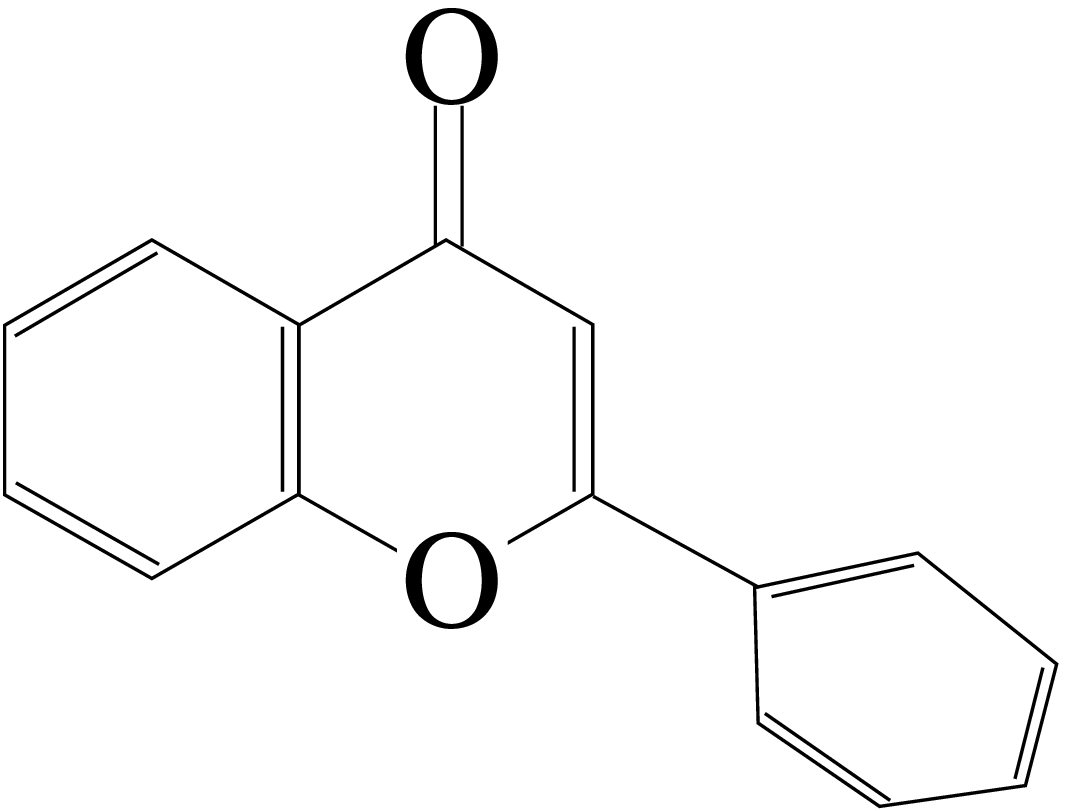}
\hspace{0.06\textwidth}
\includegraphics[width=0.24\textwidth]{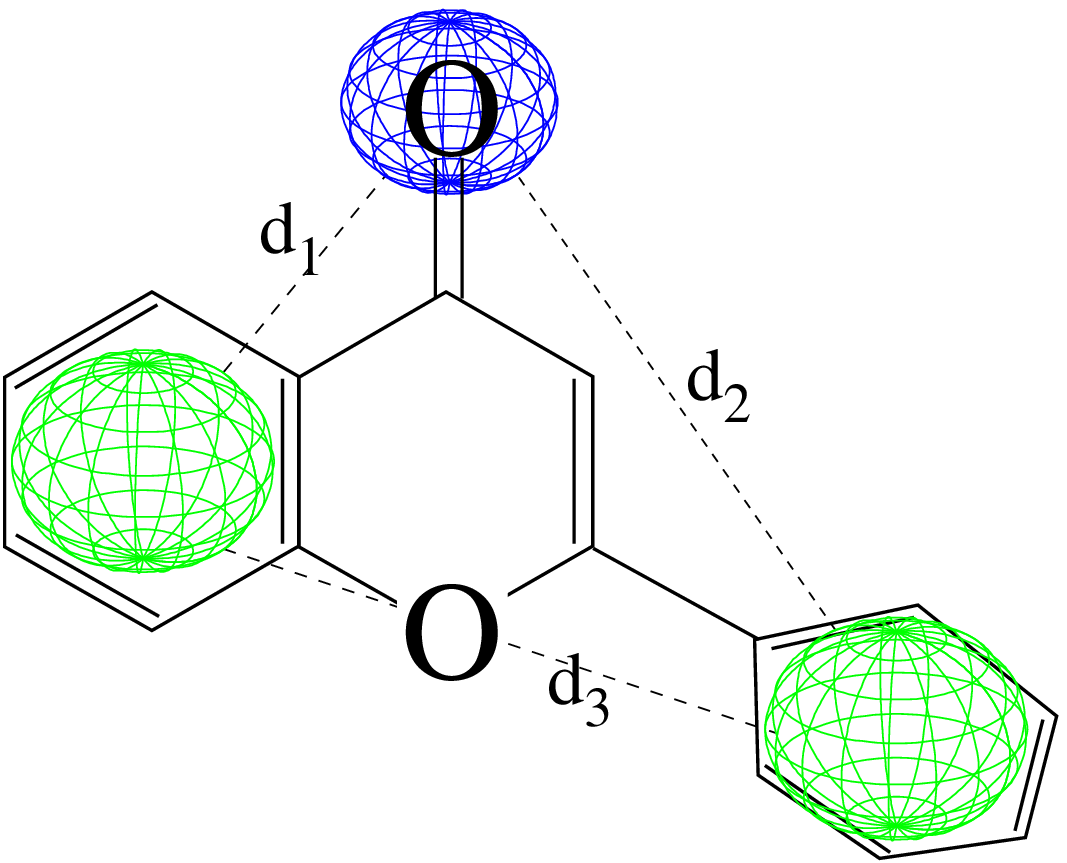}
\hspace{0.06\textwidth}
\includegraphics[width=0.24\textwidth]{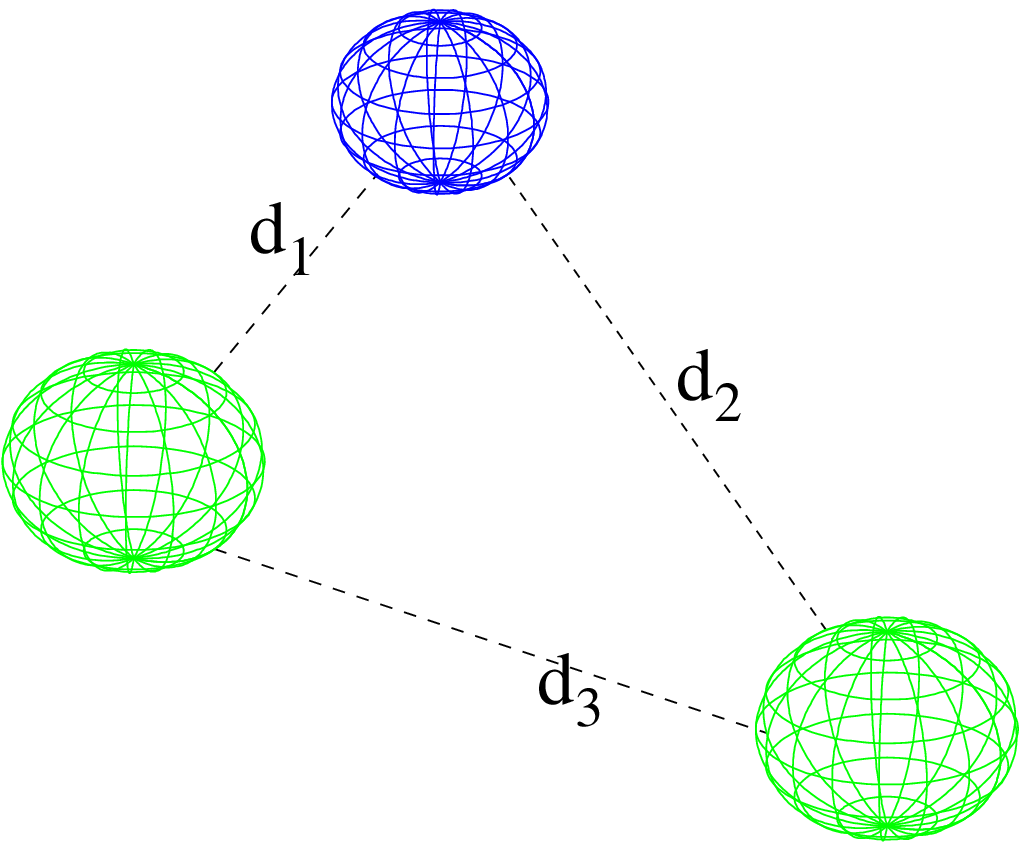}
\caption{Left: the molecule of flavone. Right: a pharmacophore made of one hydrogen bond acceptor (topmost sphere) and two aromatic rings, with distances $d_1$, $d_2$ and $d_3$ between the features that can be extracted from its structure, as shown in the middle.} \label{fig:pharmacophore}
\end{figure*}

Motivated by the fact that the tridimensional structure of molecules have a central role in many biological mechanisms, including drug-target interactions for instance, recent attempts have been made to develop kernels for 3D structure of molecules.
In this section, we introduce a class of kernels that relies on the notion of {\em pharmacophore} which is widely used in chemoinformatics.
A pharmacophore is usually defined as a spatial arrangement of three to four atoms\footnote{More generally, pharmacophore are defined as arrangements of {\em groups} of atoms having particular properties, such as positive or negative polarity, high hydrophobicity, and so on.} responsible for the biological activity of a drug molecule.
In the following, we focus on \emph{three-points pharmacophores} composed of three atoms, whose arrangement therefore forms a triangle in the 3D space (Figure~\ref{fig:pharmacophore}), but similar ideas naturally apply to pharmacophores of different cardinalities\footnote{In particular, similar ideas were developed in \citet{Swamidass2005Kernels} based on two-points pharmacophores, that is to say, distances between pairs of atoms.}.
With a slight abuse we refer as pharmacophore below to \emph{any} possible configuration of three atoms arranged as a triangle and present in a molecule, representing therefore a \emph{putative} configuration responsible for the biological property of interest. 
More precisely, we consider a molecule $m$ as a set of atoms in the 3D space, that is:  
$$
m = \cbr{ \br{x_i,l_i} \in \RR^3 \times \mc{L} }_{i=1,...,\abs{m}} \; ,
$$ where $\abs{m}$ is the number of atoms that compose the molecule, and $(x_i,l_i) \in \RR^3 \times \mc{L}$ stands for its $i$-th atom, $x_i$ being its vector of (x,y,z) coordinates, and $l_i$ its label, such as its type for instance, but more generally taken from a set $\mc{L}$ of atom labels.
With these notations at hand, the set of three-points pharmacophores that can be extracted from the molecule $m$ can be formally defined as:
\begin{equation*}
\mc{P}(m) = \cbr{ \br{p_1,p_2,p_3} \in m^3, p_1 \neq p_2 , p_1 \neq p_3, p_2 \neq p_3 } \; .
\end{equation*}

Following our discussion of Section \ref{sec:2d-kernel}, a simple way to represent a molecule $m$ is to extract all its pharmacophores, sort them by type, and count in a vector $\Phi(m)$ the number of pharmacophores of each possible type.
Clearly, the number of pharmacophores associated to a molecule is finite, but since their definition is based on the precise (x,y,z) coordinates of the atoms it is made of, or equivalently on continuous inter-atomic distances, the space of all {\it possible} pharmacophores is infinite.
Defining such a vector representation therefore requires in practice to discretize the space of pharmacophores, which boils down to discretizing the range of inter-atomic distances into a pre-defined number of bins.
Formally, if we consider $n$ bins in the discretization, this operation defines a space of {\em discrete pharmacophores} $\mc{T} = \mc{L}^3 \times [1,n]^3$, where each pharmacophore corresponds to a triplet of atom labels, taken from the alphabet $\mc{L}$, and a triplet of distance bin indices, taken in $[1,n]$.
We can now define the vector representation $\Phi(m)$, in which each coordinate $\Phi_t(m)$ is the number of pharmacophores extracted from the molecule $m$ that correspond to the discrete pharmacophore $t$, that is, 
\begin{equation*}
\Phi_t(m) = \sum_{p \in \mc{P}(m)} {\bf 1}(\text{disc}(p) = t),
\end{equation*}
where the function ${\bf 1}(\text{disc}(p) = t)$ is one if the discretized version of the pharmacophore $p$ is $t$, meaning that they are based on the same triplet of atom labels and the same triplet of distance bins, and zero otherwise.
The number $n$ of bins considered in the discretization specifies the resolution at which distinct pharmacophores are considered to be equivalent, and constitutes a critical parameterization issue. Indeed, small distance bins may prevent the detection of similar pharmacophores, while large distance bins can lead to a matching between unrelated pharmacophores.
In practice, this parameter also defines the dimension of $\Phi(m)$.
For example, considering 6 distinct types of atoms and 10 distance bins, which corresponds to bins of 2 angstroms if pairs of atoms are considered to lie within the 0-20 angstrom distance range, the cardinality of $\mc{T}$, hence the dimension of $\Phi(m)$, is $216,000$.
This number is raised up to $1,728,000$ in order to reach a precision of 1 angstrom per inter-atomic distance bin.
This explosion in the number of dimensions suggests again that in order to explicitly store the vector $\Phi(m)$, one should either consider a limited number of bins, thereby considering a poor resolution to characterize molecular structures, or rely on hashing algorithms to map the vector $\Phi(m)$ onto a vector of limited size, which as discussed previously, has the effect of inducing clashes between distinct pharmacophores.
This representation highlights once again the benefit of using kernel functions since, following the lines of Equation \ref{eq:trick}, one can define the kernel:
\begin{equation}\label{eq:pharma}
\begin{split}
k(m,m') &= \Phi(G)^\top \Phi(G') \\ 
        &= \sum_{p\in \mc{P}(m)} \sum_{p'\in \mc{P}(m')} \II(\text{disc}(p)=\text{disc}(p')),
\end{split}
\end{equation}
which, as will be discussed in Section \ref{sec:3d-implementation}, enables to map pairs of molecules and compute their inner product in feature spaces indexed by millions of pharmacophores, for a computational complexity that remains polynomial with respect to the product of their sizes.

\begin{figure}
\begin{center}
\includegraphics[width=0.3\textwidth]{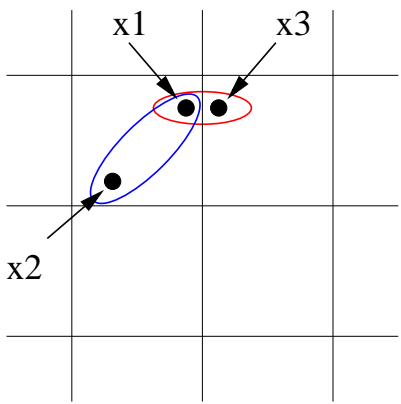}
\caption{Illustration of the discretization issue. $x_1$, $x_2$ and $x_3$ correspond to pharmacophores living in a discretized bidimensional (Euclidean) space. $x_1$ is closer to $x_3$ than it is from $x_2$, yet the discretization affects $x_1$ and $x_2$ to the same bin and $x_3$ to another bin. The kernel of Equation (\ref{eq:kernel-pharma}) allows to circumvent this issue.}\label{fig:grid}
\end{center}
\end{figure}

Of course, the idea of representing a molecule by means of its pharmacophoric content is not new, and the above approach bears strong similarity with well known pharmacophore fingerprint representations \citep{Brown1997information,Matter1999Comparing,McGregor1999Pharmacophore}.
The above discussion nevertheless illustrates the interest of using kernel functions in this case, since they allow to {\it exactly} compute the inner products between very high-dimensional feature vectors without the need of computing nor storing them, which is not possible in general and comes at the price of an information loss.
This is not, however, the major improvement made possible by kernel functions in this context.
Indeed, as shown in Figure \ref{fig:grid}, the main drawback of this approach lies in the discretization of the pharmacophore space itself: not only the choice of the discretization step controls the precision required to match a pair of pharmacophores, but it also prevents pharmacophores falling on different sides of bins edges to be matched, although they can be very close, and actually even closer that two pharmacophores falling in the same bin.
The kernel approach allows to circumvent this discretization issue by means of a simple generalization of Equation (\ref{eq:pharma}), where the binary function checking whether pairs of pharmacophores have the same discretized version or not is replaced by a general kernel between pharmacophores in order to continuously quantify their similarity.
Letting $k_P$ be such a kernel, this leads to the general 3D kernel formulation:
\begin{equation}\label{eq:kernel-pharma}
k(m,m') = \sum_{p\in \mc{P}(m)} \sum_{p'\in \mc{P}(m')} k_P(p,p'),
\end{equation}
which was introduced in \citet{Mahe2006Pharmacophore}.
A meaningful kernel $k_P$ between pharmacophores should intuitively quantify at the same the similarity of the triplets of atoms the pair of pharmacophores to be compared are defined from, and the similarity of their spatial arrangement.
A natural way to achieve this goal, which is at the same time compatible with the algorithm implementing the kernel (\ref{eq:kernel-pharma}) (see Section \ref{sec:3d-implementation}), consists in factorizing the kernel $k_P$ along the pairs of atoms and inter-atomic distances that define the pair of pharmacophores to be compared.
\citet{Mahe2006Pharmacophore} suggest for instance to introduce elementary kernel functions $k_{\text{At}}: \mc{L} \times \mc{L} \rightarrow \RR $ and $k_{\text{Dist}}: \RR \times \RR \rightarrow \RR$ comparing atoms and distances respectively, and to define the kernel $k_P$ as:
\begin{equation}\label{eq:kp}
k_P(p,p') = \prod_{i=1}^3 k_{\text{At}}(l_i,l'_i) \prod_{i=1}^3 k_{\text{Dist}}(||x_i-x_{i+1}||,||x'_{i}-x'_{i+1}||), 
\end{equation}
where the pharmacophore $p$ (resp. $p'$) is defined as $\big((l_i,x_i)\big)_{i=1:3}$ (resp. $\big((l'_i,x'_i)\big)_{i=1:3}$), $||.||$ denotes the Euclidean distance, and the index $i+1$ is taken modulo 3.
In this approach, the task of defining a kernel between 3D structures therefore boils down to defining a couple of kernels comparing atoms and inter-atomic distances.
These kernels intuitively define the elementary notions of similarity involved in the pharmacophore comparison, which in turns define the overall similarity between molecules.
A simple default choice for these kernels is to define the atom kernel $k_{\text{At}}$ as a binary kernel simply checking whether the pair of atoms to be compared have the same label or not, that is:
$$
k_{\text{At}}(l,l') = {\bf 1} (l = l'),
$$
and to define the inter-atomic distance kernel $k_{\text{Dist}}$ as the following Gaussian radial basis function (RBF) kernel:
$$
k_{\text{Dist}}(x,y) = \exp(-\frac{||x-y||^2}{\sigma^2}),
$$
where $\sigma$ is a bandwidth parameter.
Under this parameterization, it is interesting to note that the continuous kernel of Equation (\ref{eq:kernel-pharma}) and its discretized counter part of Equation (\ref{eq:pharma}) share an important feature: because the atom kernel $k_{\text{At}}$ is binary, both kernels are based on pairs of pharmacophores defined by the same triplets of atom labels.
The striking difference between the two formulations lies in the fact that in the kernel of Equation (\ref{eq:kernel-pharma}), the strength of the pharmacophore matching is {\it continuously} controlled by the parameter $\sigma$ of the (RBF) kernel comparing inter-atomic distances.
Choosing a small value of $\sigma$ corresponds to imposing a strong constraint on the spatial similarity of pharmacophores, while a larger value of $\sigma$ allows pairs of pharmacophores to be taken into account in the kernel although their spatial configurations may differ.

We conclude this section by noting that the class of kernels defined by Equation (\ref{eq:kernel-pharma}) does not have an explicit inner product interpretation in general, and in particular using the above parameterization.
Nevertheless, this construction is known to be valid as long as the kernel $k_P$ is a proper kernel function \citep{Haussler1999Convolution}.

\section{3D kernel in practice}\label{sec:3d-practice}
In this Section, we discuss general considerations related to the application of 3D kernels in practice.

        \subsection{Implementation and complexity issues}\label{sec:3d-implementation}
Without going into technical details, it can be shown that the class of pharmacophore kernels introduced in Section \ref{sec:3d} can be computed by algorithms derived from those used for the computation of 2D kernels.
Indeed, while the 3D structure of a molecule was previously defined as a set of atoms in the 3D space, it can equivalently be seen as a fully connected labeled (and undirected) graph, with atoms as vertices and inter-atomic distances as edge labels.
Under this representation, it is easy to see that computing the continuous kernel of Equation (\ref{eq:kernel-pharma}) can be interpreted as computing a walk kernel restricted to the walks that define cycles of length 3 on the graphs.
Moreover, provided the kernel $k_P$ factorizes along the pair of pharmacophore to be compared, which is the case of the kernel proposed in Equation (\ref{eq:kp}), it is easy to show that this kernel can be computed by product-graph algorithms and simple matrix product operations, for a cubic complexity with respect to the product of the sizes of the molecules to be compared.
While this complexity can be prohibitive for applications involving large datasets of molecules, the discretized version of the kernel can benefit from fast implementations derived, here also, from string kernel algorithms and trie-tree structures.
We refer the interested reader to \citet{Mahe2006Pharmacophore} for a detailed discussion about the implementation and the computational complexity of these kernels.

        \subsection{Kernel parameterization}
In its discretized version, the only parameter entering the definition of the kernel is the number of bins $n$ to discretize the inter-atomic distances.
As already noted in Section \ref{sec:3d}, this parameter is of critical importance since it controls the precision up to which pharmacophores are considered to be identical or not.
Unfortunately, this parameter can hardly be chosen a priori, and \citet{Mahe2006Pharmacophore} suggest to optimize this parameter using cross-validation procedures.
In this study, when optimized over the grid $\{4,6,8,...,30\}$ for a 0-20 angstrom inter-atomic distance range, this parameter was usually taken between 20 and 30, which suggests that the matching between a pair of pharmacophores should be subject to strong spatial constraints.
On the other hand, such fine grained resolutions have the effect of increasing the impact of the discretization issue illustrated in Figure \ref{fig:grid}.

Under the parameterization proposed in Section \ref{sec:3d}, the only parameter entering the definition of the general kernel of Equation (\ref{eq:kernel-pharma}) is the bandwidth $\sigma$ of the RBF kernel between inter-atomic distances.
In the above study, small values of $\sigma$, which correspond to strong spatial constraints in the pharmacophore comparison, are usually selected by cross-validation procedures.
While in these cases the discrete and continuous formulations of the kernel tend to coincide\footnote{Indeed, in the extreme case where $\sigma$ tends to 0 and $n$ to $+\infty$, both formulations are equivalent.}, the continuous formulation usually led to better performance in this study.

        \subsection{Molecule enrichment}
Many mechanisms of interest in tridimensional virtual screening involve specific physicochemical properties of the molecules.
In the case of drug-target interaction for instance, the molecular mechanisms responsible for the binding are known to depend on a precise 3D complementarity between the drug and the target, from both the steric and electrostatic perspectives.
For this reason, standard pharmacophore based approaches, and in particular pharmacophore fingerprints, usually define pharmacophores from atoms or groups of atoms having particular properties.
Typical molecular features of interest are  positive and negative charges, high hydrophobicity, hydrogen donors and acceptors and aromatic rings \citep{Pickett1996Diversity}.

Similarly to the introduction of Morgan indices in 2D kernels discussed in Section \ref{sec:morgan}, the atom-based kernel constructions presented in the previous section can naturally be extended to integrate this type of external information using specific label enrichment schemes.
For instance, \citet{Mahe2006Pharmacophore} use a simple scheme where the label of an atom is composed of its type and the sign of its partial charge. Positively-charged, neutral and negativaly-charged atoms of carbon are therefore labeled as $\{C^+,C^0,C^-\}$ in this approach.
Alternative labeling schemes are considered in \citet{Azencott2007One-}, based in particular on element hybridization, where for instance an {\it sp3} carbon atom is labeled as $C.3$, and a typing of atoms according to conventional pharmacophoric features, such as polarity, hydrophobicity, and hydrogen-bond donors and acceptors.
These studies show that, in general, such label enrichments have a positive influence on the subsequent structure-activity relationship models, while enabling to drastically reduce the computation cost of the kernels in some cases \citep{Mahe2006Pharmacophore}.

        \subsection{Conformational analysis}

For real-world applications, considering the tridimensional structure of molecules raises the additional issue of conformational analysis.
Indeed, because of the presence of rotational bonds, molecules are not static in the 3D space, but can alternate between several spatial configurations of low-energy called \emph{conformations}.
The mainstream approach to conformational analysis is to represent a molecule as a set of structures, called conformers, sampled from its class of admissible conformations.
On the methodological side, this operation casts the learning problem into the framework of multi-instance learning, that has been drawing a considerable interest in the machine learning community since its initial formulation \citep{Dietterich1997Solving}.
The SVM and kernel approaches lend themselves particularly well to this problem, due, on the one hand, to extensions of the SVM algorithm \citep{Andrews2002Multiple}, and, on the other hand, to the possibility to define kernels between sets of structures from a kernel between structures \citep{Gartner2002Multi-Instance}.
A possible solution to the latter approach consists in averaging kernel values over all possible pairs of conformers.
While more elaborated schemes can be adopted, such as \citep{Blaschko2006Conformal} for instance, this simple configuration was already shown to be efficient in practice by \citet{Azencott2007One-}.

\section{Discussion}\label{sec:discussion}

As a conclusion, it is probably fair to say that the empirical evaluations of the different kernel constructions introduced in this paper demonstrate the relevance of the approach based on structure kernels for virtual screening.
Indeed, on the different tasks they have been tested on, including notably the prediction of high mutagenicity molecules and drug-target inhibitors, these kernels often compare favorably to state-of-the-art approaches.
Moreover, because of the intrinsic modularity of kernel methods, this approach offers, to some extent, a unified approach to SAR and virtual screening, for two reasons.
First, because they circumvent the need of selecting and extracting molecular descriptors, these kernels can straightforwardly be used to model different biological properties.
Second, although we focused in this paper on classification applications, these kernels can be used in conjunction with the whole family of algorithms called kernel methods to solve a great variety of tasks which are relevant for virtual screening and chemoinformatics applications, such as, for instance, regression, clustering and similarity analysis.
Concerning its practical use for the screening of large datasets however, it must be stressed that the approach based on kernel methods can be computationally demanding, even for relatively small datasets. Speeding up SVM and kernel methods for large datasets is currently a topic of interest in the machine learning community, and applications in virtual screening on large databases of molecules will certainly benefit from the advances in this field.
The choice of a particular kernel, or even more importantly, of the 2D or 3D representation of molecular structures, should be dictated by the application considered.
For example, while it is widely accepted that several drug-like properties, such as intestinal absorption \citep{Lipinski2001Experimental} or mutagenicity \citep{King1996Structure-activity} for instance, can be efficiently deduced from the 2D  structure of the molecule, target binding prediction is known to depend on a precise 3D complementarity between the structures of the drug and the target, from both the steric and electrostatic perspectives \citep{Bohm2003Protein-ligand}.
Nevertheless, even in such problems that intrinsically depend on tridimensional mechanisms, it is not clear that models based on 3D kernels are more efficient than models based on 2D kernels.
This fact is especially emphasized in \citet{Azencott2007One-} where 2D kernels are shown to outperform 3D kernels\footnote{including 3D kernels based on multi-conformers.} in general, which actually tallies previous fingerprint-based studies \citep{Brown1996Use,Brown1997information}.\\

We see many potential extensions to the general kernel constructions presented above:
\begin{itemize}
        \item First, the fact that the models could benefit from simple data enrichment schemes, based, for instance, on Morgan indices in the 2D case and partial charges in the 3D case, suggests that the introduction of a more thorough chemical knowledge could improve the expressive power of the kernels.
In particular, several reduced representations of molecular structures exist, defined, for instance,  by merging aromatic cycles and atoms that are part of the same functional groups in the 2D representation \citep{Gillet2003Similarity}, or by considering generic pharmacophoric features instead of isolated atoms in the 3D case \citep{Pickett1996Diversity}.
Applying such transformations in a pre-processing step is most likely to improve the characterization of the molecular structures in the kernels, while reducing their computational cost.
        \item Other important issues that, in our opinion, would be worth studying in more details are related to conformational analysis, and more precisely to the way the conformational space of a molecule is sampled and multi-instance kernels are defined.
Although in their current form 3D kernels tend to be outperformed by their 2D counterparts \citep{Azencott2007One-}, we believe that a proper handling of multi-conformers, together with a higher level of pharmacophoric characterization of molecules, can have a great impact for virtual screening applications.
        \item Another possible extension would be to adopt a global representation of molecules and to integrate the information derived from their 1D, 2D and 3D structures. A possible approach would be to consider a single kernel defined as a linear combination of kernels for 2D and 3D structures, together with a simple kernel based on global physicochemical properties. Several methods have been proposed to optimize such a kernel combination within the framework of support-vector machines based, for instance, on semi-definite programming \citep{Lanckriet2004Kernel-based}.
        \item Finally, in the case of drug-target prediction when additional information about the structure of the target is available, it would be interesting to combine the ligand- and the structure-based approaches to virtual screening, that would most likely benefit to each other in this context.\\
\end{itemize}

Last but not least, note that this gentle introduction to kernels for molecular structures and virtual screening applications only reflects our own view and experience, and was deliberately biased towards our own developments in this field.
Indeed it must be stressed that, following the pioneer introduction of graph kernels by \citet{Kashima2003Marginalized} and \citet{Gartner2003graph}, several alternative kernel constructions have been proposed in recent years, among which:
\begin{itemize}
        \item A graph kernel based on the detection of cyclic- and tree- patterns by \citet{Horvath2004Cyclic}.
        \item A graph kernel based on the count of common paths by \citet{Borgwardt2005Shortest-Path}.
However, because it is not possible to consider exhaustive sets of paths, as mentioned in Section \ref{sec:no-totters}, the kernel construction is restricted to the sets of shortest paths between pairs of vertices.
        \item An {\em optimal assignment kernel}, based on the idea of optimally assigning the atoms from one molecule to those of another, by \citet{Frohlich2005Optimal}. This kernel formulates as the sum of a kernel between pairs of atoms, that has to be maximized over all possible assignment of the set of atoms of the smaller molecule to the set of atoms of the bigger one. Unfortunately, albeit very natural, this kernel is not positive definite and might require additional tricks to be used with kernel methods.
        \item Finally, borrowing techniques from computational geometry, standard walk-based graph kernels have recently been extended to kernels between tridimensional structures, based on graphs approximating molecular surfaces by \citet{Azencott2007One-}.
\end{itemize}
Together with the references given in the above presentation, this list constitutes, to our knowledge, a comprehensive view of kernel for molecular structures with applications in virtual screening.
As an ending remark, we would like to mention that open-source implementations of the family of kernels introduced in this paper can be found within the C++ \texttt{ChemCpp} toolbox, freely and publicly available at \texttt{http://chemcpp.sourceforge.net}.
We hope that this introductory presentation, together with the availability of this software, will help and motivate the chemoinformatics community to further investigate SVMs and molecular kernels to model structure-activity relationship.

\section*{Acknowledgments}
The work related to this paper was done while PM was PhD student in the Center for Computational Biology of the Ecole des Mines de Paris (France).
PM would like to thank Xerox Research Center Europe for letting him spend some time writing this paper.


\begin{thebibliography}{56}
\expandafter\ifx\csname natexlab\endcsname\relax\def\natexlab#1{#1}\fi
\expandafter\ifx\csname url\endcsname\relax
  \def\url#1{{\tt #1}}\fi

\bibitem[Aires-de Sousa and Gasteiger(2005)]{Aires-de-Sousa2005Prediction}
J.~Aires-de Sousa and J.~Gasteiger.
\newblock Prediction of enantiomeric excess in a combinatorial library of
  catalytic enantioselective reactions.
\newblock {\em J {C}omb {C}hem}, 7\penalty0 (2):\penalty0 298--301, 2005.
\newblock URL \url{http://dx.doi.org/10.1021/cc049961q}.

\bibitem[Andrews et~al.(2002)Andrews, Hofmann, and
  Tsochantaridis]{Andrews2002Multiple}
S.~Andrews, T.~Hofmann, and I.~Tsochantaridis.
\newblock Multiple {I}nstance {L}earning with {G}eneralized {S}upport {V}ector
  {M}achines.
\newblock In {\em Proceedings of the {E}ighteenth {N}ational {C}onference on
  {A}rtificial {I}ntelligence}, pages 943--944. American Association for
  Artificial Intelligence, 2002.

\bibitem[Arimoto et~al.(2005)Arimoto, Prasad, and
  Gifford]{Arimoto2005Development}
Rieko Arimoto, Madhu-Ashni Prasad, and Eric~M Gifford.
\newblock Development of {CYP}3{A}4 inhibition models: comparisons of
  machine-learning techniques and molecular descriptors.
\newblock {\em J {B}iomol {S}creen}, 10\penalty0 (3):\penalty0 197--205, Apr
  2005.
\newblock URL \url{http://dx.doi.org/10.1177/1087057104274091}.

\bibitem[Aronszajn(1950)]{Aronszajn1950Theory}
N.~Aronszajn.
\newblock Theory of reproducing kernels.
\newblock {\em Trans. {A}m. {M}ath. {S}oc.}, 68:\penalty0 337~--~404, 1950.

\bibitem[Azencott et~al.(2007)Azencott, Ksikes, Swamidass, Chen, Ralaivola, and
  Baldi]{Azencott2007One-}
Chloé-Agathe Azencott, Alexandre Ksikes, S.~Joshua Swamidass, Jonathan~H Chen,
  Liva Ralaivola, and Pierre Baldi.
\newblock One- to four-dimensional kernels for virtual screening and the
  prediction of physical, chemical, and biological properties.
\newblock {\em J Chem Inf Model}, 47\penalty0 (3):\penalty0 965--974, 2007.
\newblock URL \url{http://dx.doi.org/10.1021/ci600397p}.

\bibitem[Blaschko and Hofmann(2006)]{Blaschko2006Conformal}
M.B. Blaschko and T.~Hofmann.
\newblock Conformal {M}ulti-{I}nstance {K}ernels.
\newblock In NIPS 2006 {W}orkshop on {L}earning to {C}ompare {E}xamples, 2006.

\bibitem[B\"{o}hm et~al.(2003)B\"{o}hm, Schneider, Mannhold, Kubinyi, and
  Folkers]{Bohm2003Protein-ligand}
H.-J. B\"{o}hm, G.~Schneider, R.~Mannhold, H.~Kubinyi, and G.~Folkers.
\newblock {\em Protein-ligand interactions}.
\newblock Wiley, 2003.

\bibitem[Borgwardt and Kriegel(2005)]{Borgwardt2005Shortest-Path}
K.~M. Borgwardt and H.-P. Kriegel.
\newblock Shortest-path kernels on graphs.
\newblock {\em Int. {C}onf on {D}ata {M}ining}, 0:\penalty0 74--81, 2005.

\bibitem[Boser et~al.(1992)Boser, Guyon, and Vapnik]{Boser1992training}
B.~E. Boser, I.~M. Guyon, and V.~N. Vapnik.
\newblock A training algorithm for optimal margin classifiers.
\newblock In {\em Proceedings of the 5th annual {ACM} workshop on
  {C}omputational {L}earning {T}heory}, pages 144--152. ACM Press, 1992.
\newblock URL \url{http://www.clopinet.com/isabelle/Papers/colt92.ps.Z}.

\bibitem[Briem and G{\"u}nther(2005)]{Briem2005Classifying}
Hans Briem and Judith G{\"u}nther.
\newblock Classifying "kinase inhibitor-likeness" by using machine-learning
  methods.
\newblock {\em Chembiochem}, 6\penalty0 (3):\penalty0 558--66, Mar 2005.
\newblock URL \url{http://dx.doi.org/10.1002/cbic.200400109}.

\bibitem[Brown and Martin(1997)]{Brown1997information}
R.~D. Brown and Y.~C. Martin.
\newblock The information content of 2{D} and 3{D} structural descriptors
  relevant to ligand-receptor binding.
\newblock {\em J {C}hem {I}nf {C}omput {S}ci}, 37:\penalty0 1--9, 1997.

\bibitem[Brown and Martin(1996)]{Brown1996Use}
Robert~D. Brown and Yvonne~C. Martin.
\newblock Use of {S}tructure-{A}ctivity {D}ata {T}o {C}ompare
  {S}tructure-{B}ased {C}lustering {M}ethods and {D}escriptors for {U}se in
  {C}ompound {S}election.
\newblock {\em J Chem Inf Comput Sci}, 36:\penalty0 572--584, 1996.

\bibitem[Burbidge et~al.(2001)Burbidge, Trotter, Buxton, and
  Holden]{Burbidge2001Drug}
R.~Burbidge, M.~Trotter, B.~Buxton, and S.~Holden.
\newblock Drug design by machine learning: support vector machines for
  pharmaceutical data analysis.
\newblock {\em Comput. {C}hem.}, 26\penalty0 (1):\penalty0 4--15, December
  2001.
\newblock URL
  \url{http://stats.ma.ic.ac.uk/~rdb/pubs/candc-aisb00-rbmt-final.pdf}.

\bibitem[Byvatov et~al.(2003)Byvatov, Fechner, Sadowski, and
  Schneider]{Byvatov2003Comparison}
E.~Byvatov, U.~Fechner, J.~Sadowski, and G.~Schneider.
\newblock Comparison of support vector machine and artificial neural network
  systems for drug/nondrug classification.
\newblock {\em J {C}hem {I}nf {C}omput {S}ci}, 43\penalty0 (6):\penalty0
  1882--9, 2003.
\newblock URL \url{http://dx.doi.org/10.1021/ci0341161}.

\bibitem[Dietterich et~al.(1997)Dietterich, Lathrop, and
  Lozano-Perez]{Dietterich1997Solving}
T.G. Dietterich, R.H. Lathrop, and T.~Lozano-Perez.
\newblock Solving the {M}ultiple {I}nstance {P}roblem with {A}xis-{P}arallel
  {R}ectangles.
\newblock {\em Artificial Intelligence}, 89\penalty0 (1-2):\penalty0 31--71,
  1997.

\bibitem[Doniger et~al.(2002)Doniger, Hofmann, and Yeh]{Doniger2002Predicting}
S.~Doniger, T.~Hofmann, and J.~Yeh.
\newblock Predicting {CNS} permeability of drug molecules: comparison of neural
  network and support vector machine algorithms.
\newblock {\em J. {C}omput. {B}iol.}, 9\penalty0 (6):\penalty0 849--864, 2002.

\bibitem[Fr{\"o}hlich et~al.(2005)Fr{\"o}hlich, Wegner, Sieker, and
  Zell]{Frohlich2005Optimal}
H.~Fr{\"o}hlich, J.~K. Wegner, F.~Sieker, and A.~Zell.
\newblock Optimal assignment kernels for attributed molecular graphs.
\newblock In {\em Proceedings of the 22nd international conference on Machine
  learning}, pages 225 -- 232, New York, NY, USA, 2005. ACM Press.

\bibitem[G{\"a}rtner(2002)]{Gartner2002Exponential}
T.~G{\"a}rtner.
\newblock Exponential and {G}eometric {K}ernels for {G}raphs.
\newblock In NIPS {W}orkshop on {U}nreal {D}ata: {P}rinciples of {M}odeling
  {N}onvectorial {D}ata, 2002.

\bibitem[G{\"a}rtner et~al.(2003)G{\"a}rtner, Flach, and
  Wrobel]{Gartner2003graph}
T.~G{\"a}rtner, P.~Flach, and S.~Wrobel.
\newblock On graph kernels: hardness results and efficient alternatives.
\newblock In B.~Sch{\"o}lkopf and M.~Warmuth, editors, {\em Proceedings of the
  {S}ixteenth {A}nnual {C}onference on {C}omputational {L}earning {T}heory and
  the {S}eventh {A}nnual {W}orkshop on {K}ernel {M}achines}, volume 2777 of
  {\em Lecture Notes in Computer Science}, pages 129--143, Heidelberg, 2003.
\newblock URL \url{http://dx.doi.org/10.1007/b12006}.

\bibitem[G{\"a}rtner et~al.(2002)G{\"a}rtner, Flach, Kowalczyk, and
  Smola]{Gartner2002Multi-Instance}
T.~G{\"a}rtner, P.A. Flach, A.~Kowalczyk, and A.J. Smola.
\newblock Multi-{I}nstance {K}ernels.
\newblock In C.~Sammut and A.~Hoffmann, editors, {\em Proceedings of the
  {N}ineteenth {I}nternational {C}onference on {M}achine {L}earning}, pages
  179--186. Morgan Kaufmann, 2002.

\bibitem[Gasteiger and Engel(2003)]{Gasteiger2003Chemoinformatics}
J.~Gasteiger and T.~Engel, editors.
\newblock {\em Chemoinformatics : a {T}extbook}.
\newblock Wiley, 2003.

\bibitem[Gillet et~al.(2003)Gillet, Willett, and
  Bradshaw]{Gillet2003Similarity}
V.~Gillet, P.~Willett, and J.~Bradshaw.
\newblock Similarity searching using reduced graphs.
\newblock {\em J Chem Inf Comput Sci}, 43:\penalty0 338--345, 2003.

\bibitem[Hastie et~al.(2001)Hastie, Tibshirani, and
  Friedman]{Hastie2001elements}
T.~Hastie, R.~Tibshirani, and J.~Friedman.
\newblock {\em The elements of statistical learning: data mining, inference,
  and prediction}.
\newblock Springer, 2001.

\bibitem[Haussler(1999)]{Haussler1999Convolution}
D.~Haussler.
\newblock Convolution {K}ernels on {D}iscrete {S}tructures.
\newblock Technical Report UCSC-CRL-99-10, UC Santa Cruz, 1999.
\newblock URL \url{http://www.cse.ucsc.edu/~haussler/convolutions.ps}.

\bibitem[Helma et~al.(2004)Helma, Cramer, Kramer, and De~Raedt]{Helma2004Data}
C.~Helma, T.~Cramer, S.~Kramer, and L.~De~Raedt.
\newblock Data mining and machine learning techniques for the identification of
  mutagenicity inducing substructures and structure activity relationships of
  noncongeneric compounds.
\newblock {\em J {C}hem {I}nf {C}omput {S}ci}, 44\penalty0 (4):\penalty0
  1402--11, 2004.
\newblock URL \url{http://dx.doi.org/10.1021/ci034254q}.

\bibitem[Horv{\'a}th et~al.(2004)Horv{\'a}th, G{\"a}rtner, and
  Wrobel]{Horvath2004Cyclic}
T.~Horv{\'a}th, T.~G{\"a}rtner, and S.~Wrobel.
\newblock Cyclic pattern kernels for predictive graph mining.
\newblock In {\em Proceedings of the tenth ACM SIGKDD international conference
  on Knowledge discovery and data mining}, pages 158--167, New York, NY, USA,
  2004. ACM Press.

\bibitem[Kashima et~al.(2003)Kashima, Tsuda, and
  Inokuchi]{Kashima2003Marginalized}
H.~Kashima, K.~Tsuda, and A.~Inokuchi.
\newblock Marginalized {K}ernels between {L}abeled {G}raphs.
\newblock In T.~Faucett and N.~Mishra, editors, {\em Proceedings of the
  {T}wentieth {I}nternational {C}onference on {M}achine {L}earning}, pages
  321--328. AAAI Press, 2003.

\bibitem[Kashima et~al.(2004)Kashima, Tsuda, and Inokuchi]{Kashima2004Kernels}
H.~Kashima, K.~Tsuda, and A.~Inokuchi.
\newblock Kernels for graphs.
\newblock In B.~Sch{\"o}lkopf, K.~Tsuda, and J.P. Vert, editors, {\em Kernel
  {M}ethods in {C}omputational {B}iology}, pages 155--170. MIT Press, 2004.

\bibitem[King et~al.(1996)King, Muggleton, Srinivasan, and
  Sternberg]{King1996Structure-activity}
R.~D. King, S.~H. Muggleton, A.~Srinivasan, and M.~J. Sternberg.
\newblock {S}tructure-activity relationships derived by machine learning: the
  use of atoms and their bond connectivities to predict mutagenicity by
  inductive logic programming.
\newblock {\em Proc Natl Acad Sci U S A}, 93\penalty0 (1):\penalty0 438--442,
  Jan 1996.

\bibitem[Kramer et~al.(2002)Kramer, Frank, and Helma]{Kramer2002Fragment}
S.~Kramer, E.~Frank, and C.~Helma.
\newblock Fragment generation and support vector machines for inducing {SAR}s.
\newblock {\em S{AR} {QSAR} {E}nviron {R}es}, 13\penalty0 (5):\penalty0
  509--23, Jul 2002.
\newblock URL \url{http://dx.doi.org/10.1080/10629360290023340}.

\bibitem[Lanckriet et~al.(2004)Lanckriet, Cristianini, Jordan, and
  Noble]{Lanckriet2004Kernel-based}
G.R.G. Lanckriet, N.~Cristianini, M.I. Jordan, and W.S. Noble.
\newblock Kernel-based integration of genomic data using semidefinite
  programming.
\newblock In B.~Schölkopf, K.~Tsuda, and J.P. Vert, editors, {\em Kernel
  {M}ethods in {C}omputational {B}iology}, pages 231--259. MIT Press, 2004.

\bibitem[Leslie et~al.(2002)Leslie, Eskin, and Noble]{Leslie2002Spectrum}
C.~Leslie, E.~Eskin, and W.S. Noble.
\newblock The spectrum kernel: a string kernel for {SVM} protein
  classification.
\newblock In Russ~B. Altman, A.~Keith Dunker, Lawrence Hunter, Kevin Lauerdale,
  and Teri~E. Klein, editors, {\em Proceedings of the {P}acific {S}ymposium on
  {B}iocomputing 2002}, pages 564--575. World Scientific, 2002.
\newblock URL
  \url{http://www.smi.stanford.edu/projects/helix/psb02/leslie.pdf}.

\bibitem[Lind and Maltseva(2003)]{Lind2003Support}
P.~Lind and T.~Maltseva.
\newblock Support vector machines for the estimation of aqueous solubility.
\newblock {\em J {C}hem {I}nf {C}omput {S}ci}, 43\penalty0 (6):\penalty0
  1855--9, 2003.
\newblock URL \url{http://dx.doi.org/10.1021/ci034107s}.

\bibitem[Lipinski et~al.(2001)Lipinski, Lombardo, Dominy, and
  Feeney]{Lipinski2001Experimental}
C.~A. Lipinski, F.~Lombardo, B.~W. Dominy, and P.~J. Feeney.
\newblock Experimental and computational approaches to estimate solubility and
  permeability in drug discovery and development settings.
\newblock {\em Adv. {D}rug. {D}eliv. {R}ev}, 46\penalty0 (1-3):\penalty0 3--26,
  Mar 2001.

\bibitem[Liu et~al.(2004{\natexlab{a}})Liu, Zhang, Yao, Liu, Hu, and
  Fan]{Liu2004Prediction}
H.~X. Liu, R.~S. Zhang, X.~J. Yao, M.~C. Liu, Z.~D. Hu, and B.~T. Fan.
\newblock Prediction of the isoelectric point of an amino acid based on
  {GA}-{PLS} and {SVM}s.
\newblock {\em J {C}hem {I}nf {C}omput {S}ci}, 44\penalty0 (1):\penalty0
  161--7, 2004{\natexlab{a}}.
\newblock URL \url{http://dx.doi.org/10.1021/ci034173u}.

\bibitem[Liu et~al.(2004{\natexlab{b}})Liu, Zhang, Yao, Liu, Hu, and
  Fan]{Liu2004QSAR}
H.~X. Liu, R.~S. Zhang, X.~J. Yao, M.~C. Liu, Z.~D. Hu, and B.~T. Fan.
\newblock Q{SAR} and classification models of a novel series of {COX}-2
  selective inhibitors: 1,5-diarylimidazoles based on support vector machines.
\newblock {\em J {C}omput {A}ided {M}ol {D}es}, 18\penalty0 (6):\penalty0
  389--99, Jun 2004{\natexlab{b}}.

\bibitem[Luan et~al.(2005)Luan, Zhang, Zhao, Yao, Liu, Hu, and
  Fan]{Luan2005Classification}
Feng Luan, Ruisheng Zhang, Chunyan Zhao, Xiaojun Yao, Mancang Liu, Zhide Hu,
  and Botao Fan.
\newblock Classification of the carcinogenicity of {N}-nitroso compounds based
  on support vector machines and linear discriminant analysis.
\newblock {\em Chem {R}es {T}oxicol}, 18\penalty0 (2):\penalty0 198--203, Feb
  2005.
\newblock URL \url{http://dx.doi.org/10.1021/tx049782q}.

\bibitem[Mah{\'e} et~al.(2006)Mah{\'e}, Ralaivola, Stoven, and
  Vert]{Mahe2006Pharmacophore}
P.~Mah{\'e}, L.~Ralaivola, V.~Stoven, and J.-P. Vert.
\newblock The pharmacophore kernel for virtual screening with support vector
  machines.
\newblock {\em J Chem Inf Model}, 46\penalty0 (5):\penalty0 2003--2014, 2006.
\newblock URL \url{http://dx.doi.org/10.1021/ci060138m}.

\bibitem[Mah{\'e} et~al.(2005)Mah{\'e}, Ueda, Akutsu, Perret, and
  Vert]{Mahe2005Graph}
P.~Mah{\'e}, N.~Ueda, T.~Akutsu, J.-L. Perret, and J.-P. Vert.
\newblock Graph kernels for molecular structure-activity relationship analysis
  with support vector machines.
\newblock {\em J {C}hem {I}nf {M}odel}, 45\penalty0 (4):\penalty0 939--51,
  2005.
\newblock URL \url{http://dx.doi.org/10.1021/ci050039t}.

\bibitem[Mah{\'e} and Vert(2006)]{Mahe2006Graph}
P.~Mah{\'e} and J.-P. Vert.
\newblock Graph kernels based on tree patterns for molecules.
\newblock Technical Report ccsd-00095488, HAL, September 2006.
\newblock URL \url{https://hal.ccsd.cnrs.fr/ccsd-00095488}.

\bibitem[Matter and P\"{o}tter(1999)]{Matter1999Comparing}
H.~Matter and T.~P\"{o}tter.
\newblock Comparing 3{D} pharmacophore triplets and 2{D} fingerprints for
  selecting diverse compound subsets.
\newblock {\em J {C}hem {I}nf {C}omput {S}ci}, 39:\penalty0 1211--1225, 1999.

\bibitem[McGregor and Muskal(1999)]{McGregor1999Pharmacophore}
M.~J. McGregor and S.~M. Muskal.
\newblock {P}harmacophore fingerprinting. 1. {A}pplication to {QSAR} and
  focused library design.
\newblock {\em J Chem Inf Comput Sci}, 39\penalty0 (3):\penalty0 569--574,
  1999.

\bibitem[M{\"u}ller et~al.(2005)M{\"u}ller, R{\"a}tsch, Sonnenburg, Mika,
  Grimm, and Heinrich]{Mueller2005Classifying}
K.-R. M{\"u}ller, G.~R{\"a}tsch, S.~Sonnenburg, S.~Mika, M.~Grimm, and
  N.~Heinrich.
\newblock Classifying 'drug-likeness' with {K}ernel-based learning methods.
\newblock {\em J {C}hem {I}nf {M}odel}, 45\penalty0 (2):\penalty0 249--53,
  2005.
\newblock URL \url{http://dx.doi.org/10.1021/ci049737o}.

\bibitem[Pickett et~al.(1996)Pickett, Mason, and McLay]{Pickett1996Diversity}
S.~D. Pickett, J.~S. Mason, and I.~M. McLay.
\newblock Diversity profiling and design using 3{D} pharmacophores :
  {P}harmacophores-{D}erived {Q}ueries ({PQD}).
\newblock {\em J {C}hem {I}nf {C}omput {S}ci}, 36:\penalty0 1214--1223, 1996.

\bibitem[Ralaivola et~al.(2005)Ralaivola, Swamidass, Saigo, and
  Baldi]{Ralaivola2005Graph}
L.~Ralaivola, S.~J. Swamidass, H.~Saigo, and P.~Baldi.
\newblock Graph kernels for chemical informatics.
\newblock {\em Neural {N}etw.}, 18\penalty0 (8):\penalty0 1093--1110, Sep 2005.
\newblock URL \url{http://dx.doi.org/10.1016/j.neunet.2005.07.009}.

\bibitem[Ramon and G\"{a}rtner(2003)]{Ramon2003Expressivity}
J.~Ramon and T.~G\"{a}rtner.
\newblock {E}xpressivity versus efficiency of graph kernels.
\newblock In T.~Washio and L.~De~Raedt, editors, {\em {P}roceedings of the
  {F}irst {I}nternational {W}orkshop on {M}ining {G}raphs, {T}rees and
  {S}equences}, pages 65--74, 2003.

\bibitem[Saeh et~al.(2005)Saeh, Lyne, Takasaki, and Cosgrove]{Saeh2005Lead}
J.~Saeh, P.~Lyne, B.~Takasaki, and D.~Cosgrove.
\newblock Lead hopping using {SVM} and 3{D} pharmacophore fingerprints.
\newblock {\em J {C}hem {I}nf {M}odel}, 45\penalty0 (4):\penalty0 1122--1133,
  Jul 2005.
\newblock URL \url{http://dx.doi.org/10.1021/ci049732r}.

\bibitem[Sch{\"o}lkopf and Smola(2002)]{Scholkopf2002Learning}
B.~Sch{\"o}lkopf and A.~J. Smola.
\newblock {\em Learning with {K}ernels: {S}upport {V}ector {M}achines,
  {R}egularization, {O}ptimization, and {B}eyond}.
\newblock MIT Press, Cambridge, MA, 2002.
\newblock URL \url{http://www.learning-with-kernels.org}.

\bibitem[Sch{\"o}lkopf et~al.(2004)Sch{\"o}lkopf, Tsuda, and
  Vert]{Schoelkopf2004Kernel}
B.~Sch{\"o}lkopf, K.~Tsuda, and J.-P. Vert.
\newblock {\em Kernel {M}ethods in {C}omputational {B}iology}.
\newblock MIT Press, 2004.

\bibitem[Shawe-Taylor and Cristianini(2004)]{Shawe-Taylor2004Kernel}
J.~Shawe-Taylor and N.~Cristianini.
\newblock {\em Kernel {M}ethods for {P}attern {A}nalysis}.
\newblock Cambridge University Press, 2004.

\bibitem[Swamidass et~al.(2005)Swamidass, Chen, Bruand, Phung, Ralaivola, and
  Baldi]{Swamidass2005Kernels}
S.~J. Swamidass, J.~Chen, J.~Bruand, P.~Phung, L.~Ralaivola, and P.~Baldi.
\newblock Kernels for small molecules and the prediction of mutagenicity,
  toxicity and anti-cancer activity.
\newblock {\em Bioinformatics}, 21\penalty0 (Suppl. 1):\penalty0 i359--i368,
  Jun 2005.
\newblock URL \url{http://dx.doi.org/10.1093/bioinformatics/bti1055}.

\bibitem[Takaoka et~al.(2003)Takaoka, Endo, Yamanobe, Kakinuma, Okubo,
  Shimazaki, Ota, Sumiya, and Yoshikawa]{Takaoka2003Development}
Y.~Takaoka, Y.~Endo, S.~Yamanobe, H.~Kakinuma, T.~Okubo, Y.~Shimazaki, T.~Ota,
  S.~Sumiya, and K.~Yoshikawa.
\newblock Development of a method for evaluating drug-likeness and ease of
  synthesis using a data set in which compounds are assigned scores based on
  chemists' intuition.
\newblock {\em J {C}hem {I}nf {C}omput {S}ci}, 43\penalty0 (4):\penalty0
  1269--75, 2003.
\newblock URL \url{http://dx.doi.org/10.1021/ci034043l}.

\bibitem[Tobita et~al.(2005)Tobita, Nishikawa, and
  Nagashima]{Tobita2005discriminant}
M.~Tobita, T.~Nishikawa, and R.~Nagashima.
\newblock A discriminant model constructed by the support vector machine method
  for {HERG} potassium channel inhibitors.
\newblock {\em Bioorg. {M}ed. {C}hem. {L}ett.}, 15\penalty0 (11):\penalty0
  2886--90, Jun 2005.
\newblock URL \url{http://dx.doi.org/10.1016/j.bmcl.2005.03.080}.

\bibitem[Vapnik(1998)]{Vapnik1998Statistical}
V.~N. Vapnik.
\newblock {\em Statistical {L}earning {T}heory}.
\newblock Wiley, New-York, 1998.

\bibitem[Vishwanathan et~al.(2007)Vishwanathan, Borgwardt, and
  Schraudolph]{Vishwanathan2007Fast}
S.V.N. Vishwanathan, K.~Borgwardt, and N.~Schraudolph.
\newblock Fast {C}omputation of {G}raph {K}ernels.
\newblock In B.~Sch\"{o}lkopf, J.~Platt, and T.~Hoffman, editors, {\em Adv.
  {N}eural {I}nform. {P}rocess. {S}yst.}, volume~19, pages 1--2, Cambridge, MA,
  2007. MIT Press, Cambridge, MA.

\bibitem[Weston et~al.(2003)Weston, P{\'e}rez-Cruz, Bousquet, Chapelle,
  Elisseeff, and Sch{\"o}lkopf]{Weston2003Feature}
J.~Weston, F.~P{\'e}rez-Cruz, O.~Bousquet, O.~Chapelle, A.~Elisseeff, and
  B.~Sch{\"o}lkopf.
\newblock Feature selection and transduction for prediction of molecular
  bioactivity for drug design.
\newblock {\em Bioinformatics}, 19\penalty0 (6):\penalty0 764--771, 2003.
\newblock URL
  \url{http://bioinformatics.oupjournals.org/cgi/content/abstract/19/6/764}.

\end{thebibliography}
\end{document}